\documentclass[journal,onecolumn,10pt,draftclsnofoot]{IEEEtran}
\usepackage[T1]{fontenc}
\usepackage{cite}
\usepackage{amsmath,amssymb}
\usepackage{algorithm}
\usepackage{graphicx}
\usepackage{textcomp}
\usepackage{xcolor}
\usepackage{algorithm} 
\usepackage{algpseudocode} 
\usepackage{dsfont}
\usepackage{changepage}
\usepackage{setspace}
\usepackage{pdfpages}
\usepackage{mathrsfs}
\usepackage{diagbox}
\usepackage{multirow}
\usepackage{lscape}
\usepackage{rotating}
\usepackage{subfig}
\usepackage{lipsum}
\usepackage[normalem]{ulem}
\usepackage{booktabs}

\newtheorem{theorem}{Theorem}
\newtheorem{remark}{Remark}

\newtheorem{Definition}{Definition}
\newtheorem{example}{Example}
\newtheorem{lemma}{Lemma}

\begin{document}

\title{Task-Oriented Lossy Compression with Data, Perception, and Classification Constraints}

\author{
  Yuhan Wang,~\IEEEmembership{Student Member,~IEEE,} Youlong Wu,~\IEEEmembership{Member,~IEEE,}
  Shuai Ma,~\IEEEmembership{Member,~IEEE,}\\
  and Ying-Jun Angela Zhang,~\IEEEmembership{Fellow,~IEEE}
  \thanks{This work was supported in part by the General Research Fund (project number 14202421, 14214122, 14202723), Area of Excellence Scheme grant (project number AoE/E-601/22-R), and NSFC/RGC Collaborative Research Scheme (project number CRS\_HKUST603/22), all from the Research Grants Council of Hong Kong. The work of Youlong Wu was supported in part by the Platform of Computer and Communication provided by ShanghaiTech University. The work of Shuai Ma was supported in part by the  National Natural Science Foundation of China (NSFC) under Grant 62471270, and in part by Guangdong Basic and Applied Basic Research Foundation under Grant  2024A1515030028. This paper was in part presented at the IEEE Information Theory Workshop (ITW), Shenzhen, China, Nov. 2024\cite{ITW24RDPC}. \emph{(Corresponding author: Youlong Wu.)}}
  \thanks{Yuhan Wang and Ying-Jun Angela Zhang are with the Department of Information Engineering, The Chinese University of Hong Kong, Hong Kong (e-mail: wy023@ie.cuhk.edu.hk; yjzhang@ie.cuhk.edu.hk).}
  \thanks{Youlong Wu is with the School of Information Science and Technology, ShanghaiTech University, Shanghai 201210,
  China (e-mail: wuyl1@shanghaitech.edu.cn).}
  \thanks{Shuai Ma is with the Peng Cheng Laboratory, Shenzhen 518055, China (e-mail: mash01@pcl.ac.cn).}}



\maketitle

  \begin{abstract}
    By extracting task-relevant information while maximally compressing the input, the information bottleneck (IB) principle has provided a guideline for learning effective and robust representations of the target inference. However, extending the idea to the multi-task learning scenario with joint consideration of generative tasks and traditional reconstruction tasks remains unexplored. This paper addresses this gap by reconsidering the lossy compression problem with diverse constraints on data reconstruction, perceptual quality, and classification accuracy. Firstly, we study two ternary relationships, namely, the \emph{rate-distortion-classification (RDC)} and \emph{rate-perception-classification (RPC)}. For both RDC and RPC functions, we derive the \emph{closed-form expressions} of the optimal rate for binary and Gaussian sources. These new results complement the IB principle and provide insights into effectively extracting task-oriented information to fulfill diverse objectives.
    Secondly, unlike prior research demonstrating a tradeoff between classification and perception in signal restoration problems,  we prove that such a tradeoff does not exist in the RPC function and reveal that the source noise plays a decisive role in the classification-perception tradeoff.
    Finally, we implement a deep-learning-based image compression framework, incorporating multiple tasks related to distortion, perception, and classification. The experimental results coincide with the theoretical analysis and verify the effectiveness of our generalized IB in balancing various task objectives.
\end{abstract}


\section{Introduction}

\IEEEPARstart{T}{he} last decades have witnessed significant achievements in machine learning, particularly with the success of deep learning (DL) methods across various tasks. Leveraging information-theoretical insights can facilitate the understanding of DL algorithms and aid in designing loss functions. The information bottleneck (IB) principle\cite{IB} emerged as a promising theory for analyzing and training DL algorithms, especially in the context of feature extraction. The IB principle aims to find a compressed mapping $\hat X$ of the input observation $X$ while preserving enough information about a correlated target $S$. Specifically, the IB problem is given by
\begin{align*}
  &\min_{p_{\hat{X}|X}:I(S;\hat{X}) \geq C} I(X;\hat{X}),
\end{align*}
where $C$ is a constant and $S-X-\hat X$ forms Markov chain. The IB principle  can be viewed as a remote lossy source-coding problem with logarithmic loss distortion\cite{logarithmic_loss,IB_distortion}. 
Since its introduction in 1999, the IB principle has received significant attention and has been widely applied to design machine learning (e.g., representation learning) \cite{IBappML,OpenBx}  and  task-oriented communications (e.g., edge inference) \cite{TaskOriented,RobustIB_TaskOriented_Digital_Xie2023}. 

On the other hand, recent research has introduced the concept of perception quality to the lossy compression problem. The authors in \cite{RethinkingRDP} proposed the rate-distortion-perception (RDP) function, and analyzed this ternary tradeoff. Here the perception is defined to be the divergence between the source and reconstruction distributions. In \cite{rate_distortion_perception_conditional_percep}, the perception measure based on the divergence between distributions conditioned on the encoder output is also studied, which empirically results in higher perceptual quality of reconstructions\cite{mentzer2020HiFiC}. In \cite{UniversalRDPs}, the closed-form expression of the RDP function for the Gaussian source is derived under mean-square error (MSE) distortion and squared Wasserstein-2 distance. Notably, when considering perfect realism where the perception loss is zero, the RDP function aligns with the theory of distribution-preserving rate-distortion function studied in \cite{Distribution_Preserving_Quantization_Li11}. The cost of perfect realism in lossy compression was explored in \cite{2021PerceptualReconstruction}, and the advantage of stochastic encoders in the one-shot setting was illustrated in \cite{theis2021advantages_stochastic}. Additionally, different coding schemes for the RDP tradeoff have been established recently \cite{CodingTheorem_RDP_Theis2021,OnRDP_Coding_Chen2022}, demonstrating that the RDP function can be achieved by stochastic or deterministic codes.

\begin{table}[t]
  \centering
  \caption{Summary of tradeoffs between distortion, perception and classification tasks in lossy compression and signal restoration.}\label{Table_Summary_tradeoffs}
  \renewcommand\arraystretch{1.1}
  \begin{tabular}{|ccc|c|}
  \toprule
  \hline             
  \multicolumn{1}{|c|}{Ref.} & 
  \multicolumn{1}{c|}{Tradeoff} & 
  Closed form & 
  Note \\ 
  \hline
  \multicolumn{1}{|c|}{\multirow{2}{*}{\cite{cover2012elements}}} & 
  \multicolumn{1}{c|}{\multirow{2}{*}{RD}} & 
  Binary\cite{cover2012elements} & 
  \multirow{14}{*}{
    \begin{tabular}[c]{@{}l@{}} 
      Source noise\\ 
      in signal \\ 
      restoration\\
      makes the \\
      perfect\\ 
      restoration\\ 
      impossible,\\
      leading to\\ 
      persistence \\
      of tradeoffs \\
      even without \\
      a rate\\ 
      constraint.\\ 
      The details\\
      will be \\ 
      discussed in \\
      Section \ref{SecDistortion}.
    \end{tabular}} \\ 
  \cline{3-3}
  \multicolumn{1}{|c|}{} & 
  \multicolumn{1}{c|}{} & 
  Gaussian\cite{cover2012elements} &  \\ 
  \cline{1-3}
  \multicolumn{1}{|c|}{\multirow{2}{*}{\cite{IB}}} & 
  \multicolumn{1}{c|}{\multirow{2}{*}{RC}} & 
  Binary\cite{MrsGerbers} &  \\ 
  \cline{3-3}
  \multicolumn{1}{|c|}{} & 
  \multicolumn{1}{c|}{} &
  Gaussian\cite{IBGaussian} &  \\ 
  \cline{1-3}
  \multicolumn{1}{|c|}{\multirow{2}{*}{\cite{RethinkingRDP}}} & 
  \multicolumn{1}{c|}{\multirow{2}{*}{RDP}} & 
  Binary\cite{RethinkingRDP} &  \\ 
  \cline{3-3}
  \multicolumn{1}{|c|}{} & 
  \multicolumn{1}{c|}{} &
  Gaussian\cite{UniversalRDPs} &  \\ 
  \cline{1-3}
  \multicolumn{1}{|c|}{\multirow{1}{*}{\cite{PD-tradeoff}}} & 
  \multicolumn{1}{c|}{DP} & 
  Gaussian\cite{PD-tradeoff} &  \\ 
  \cline{1-3}
  \multicolumn{1}{|c|}{\multirow{1}{*}{\cite{CDP}}} & 
  \multicolumn{1}{c|}{CDP} & 
  \begin{tabular}[c]{@{}l@{}}Gaussian mixture\\ (Simulation)\cite{CDP}\end{tabular} &  \\ 
  \cline{1-3}
  \multicolumn{1}{|c|}{\multirow{5}{*}{Ours}} & 
  \multicolumn{1}{c|}{\multirow{2}{*}{RDC}} & 
  Binary &  \\ 
  \cline{3-3}
  \multicolumn{1}{|c|}{} & 
  \multicolumn{1}{c|}{} & 
  Gaussian &  \\ 
  \cline{2-3}
  \multicolumn{1}{|c|}{} & 
  \multicolumn{1}{c|}{\multirow{2}{*}{RPC}} & 
  Binary &  \\
  \cline{3-3}
  \multicolumn{1}{|c|}{} & 
  \multicolumn{1}{c|}{} &
  Gaussian &  \\ 
  \cline{2-3}
  \multicolumn{1}{|c|}{} & 
  \multicolumn{1}{c|}{\begin{tabular}[c]{@{}c@{}}RPC\\ given D\end{tabular}} & 
  \begin{tabular}[c]{@{}c@{}}Gaussian\\ (Simulation)\end{tabular} &  \\ 
  \cline{1-3}
  \multicolumn{3}{|c|}{\begin{tabular}[c]{@{}l@{}}\textbf{Note:} R, D, P, and C refer to rate, data \\distortion, perception and classification.\end{tabular}} &  \\ 
  \hline
  \end{tabular}
  \vspace{-0.4cm}
  \end{table}

  Similar tradeoffs have also been explored in the problem of signal degradation and restoration, where signals are corrupted by extrinsic noise and only a degraded version of the source signal is available for reconstruction or denoising.
According to the study in \cite{PD-tradeoff}\cite{DP-tradeoff_Wasserstein_Freirich2021}, there is not always a direct correlation between the technical reduction of distortion and the enhancement of perceived visual quality. 
The authors mathematically proved the existence of a tradeoff between distortion and perception, known as the distortion-perception (DP) tradeoff. 
Furthermore, the consideration of classification tasks alongside perceptual quality and distortion was addressed in \cite{CDP}. The authors investigated the tradeoff between classification, distortion, and perception, referred to as the classification-distortion-perception (CDP) tradeoff. The experimental results highlight that achieving better classification performance often comes at the expense of higher distortion or poorer perceptual quality.

\subsection{Motivations}
Although existing works provide certain explanations and design guidance for DL in various tasks, theoretical analysis is still missing for many important applications. 
For example,  the development of smart cities and the Internet of Things (IoT) pose new challenges in efficiently compressing and analyzing massive data to meet diverse objectives, including pixel-level reconstruction and feature extraction\cite{VCM_TIP2020,VCM_TPAMI2024}. In particular, distortion is concerned jointly with classification accuracies in scenarios such as autonomous driving and face recognition in surveillance systems\cite{ImageQuality_DNN_Samuel2016, 2C-Net_ImageCompress_Classification}. Meanwhile, perceptual quality and fidelity on label information are often jointly considered in the context of conditional generative adversarial nets (C-GAN)\cite{conditionalGAN_mirza2014}, InfoGAN\cite{InfoGAN}, and CatGAN\cite{CatGAN}. The loss functions are usually a composition of GAN-loss and mutual information regularization. In these scenarios, there is a lack of sufficient theoretical characterization of the relationships between target inference tasks and generative tasks as well as traditional reconstruction tasks.

Furthermore, multi-task machine learning is more complex compared to single-task learning. In particular, different tasks may exhibit conflicting needs, making it crucial to explore the tradeoffs among them under limited resources\cite{Multi-Task_learning_Survey}. Meanwhile, although several heuristic designs for multi-task loss functions have been proposed\cite{Multi-Task_learning_Survey}, no guidelines on balancing different tasks have been provided from an information-theoretical perspective. The existing IB principle has been successfully applied to single-task scenarios with a focus on feature extraction. A multi-task information bottleneck theory with joint consideration of generative tasks and even traditional restoration tasks is more desirable. Such a theory can provide insights into understanding tradeoffs in diverse tasks, which has wide applications in next-generation task-oriented communication\cite{Wu_Effective_Semcom,Shi_Semcom_Maga}. Meanwhile, it has great potential to offer design guidelines for multi-task learning.

\subsection{Our Contributions}
 To address the above issues, we integrate traditional reconstruction tasks and generative tasks into the IB principle and investigate both the theoretical and practical roles of the generalized IB framework. Our contributions are summarized as follows.
\begin{itemize}
  \item First, we investigate two open problems: the rate-distortion-classification (RDC) function and the rate-perception-classification (RPC) function. We derive the \emph{closed-form expressions} for both RDC and RPC functions in binary and scalar Gaussian cases. Notably, both RDC and RPC functions exhibit distinct characteristics compared to previous theories. They complement IB theory and RDP tradeoff, and provide new insights and frameworks on characterizing task-relevant information.
  \item    In contrast to existing work showing the existence of a tradeoff between classification and perception \cite{CDP} in the signal degradation and restoration problem, our theoretical analysis shows that such a tradeoff does not exist in the RPC function. To account for this contradiction, we reveal that the \emph{source noise} plays a decisive role, and show that once given a certain level of distortion, the classification-perception tradeoff appears in the RPC function. 
  \item   Finally, we conduct a series of experiments by implementing a DL-based image compression framework incorporating multiple tasks. The experimental outcomes validate our theoretical results on the RDC, RPC tradeoffs, as well as RPC with certain levels of distortion.  The experiments demonstrate that the generalized IB framework could effectively provide guidelines for designing DL learning methods with multiple task objectives.
\end{itemize}

Table \ref{Table_Summary_tradeoffs} provides a summary of the tradeoffs between distortion, perception, and classification tasks in both lossy compression and signal restoration, as presented in previous literature and this paper. The tradeoff between rate and classification (RC) refers to the IB principle\cite{IB}.  Note that although the CDP tradeoff has been previously studied in \cite{CDP} in the view of signal restoration, our settings differ greatly in two aspects: First, the tradeoffs between different tasks are evaluated in compressed scenarios, which could affect the behavior of all three tasks. Second, the analysis of classification performance in \cite{CDP} is characterized by a predefined classifier while we use conditional entropy with broader applications in feature extraction and target inference.

The rest of this paper is organized as follows. In Section \ref{SecProblemFormulation} we present the problem formulation and the proposed information rate-distortion-perception-classification function. The RDC and RPC functions are investigated in Section \ref{SecRDC-RPC}. In Section \ref{SecDistortion}, we discuss the decisive role of source noise in the existence of tradeoffs. Then, experimental results are presented in Section \ref{SecExperimentResults}. Finally, Section \ref{SecConclusion} concludes the paper.

\textbf{Notations:} For a random variable $X$ denoted as a capital letter, we use small letter $x$ for its realizations, and use $p_X(x)$ to denote the distribution over its alphabet $\mathcal{X}$. The expectation is denoted as $\mathbb E[X]$. We use $H(\cdot)$ to denote the Shannon entropy of a discrete random variable, and $h(\cdot)$ to represent the differential entropy.

\section{Problem Formulation} \label{SecProblemFormulation}
Consider a source with observable data $X\sim p_X(x)$, which intrinsically contains several target labels formulated by variables $S_1,\cdots,S_K\sim p_S(s_1,\cdots,s_K)$. The observation $X$ and the intrinsic variables are correlated and follow a joint probability distribution $p_{X,S}(x,s_1,\cdots,s_K)$ over $\mathcal{X}\times\mathcal{S}_1\times\cdots\times\mathcal{S}_K$. The label variables are not observable but could be inferred from $X$. For example, the observation $X$ could be a voice signal, and the interested classification variables may be the transcription of the speech, the identity of the speaker, or the gender of the speaker. 

\begin{figure}[!htbp]
  \centering
    \includegraphics[width=0.56\textwidth]{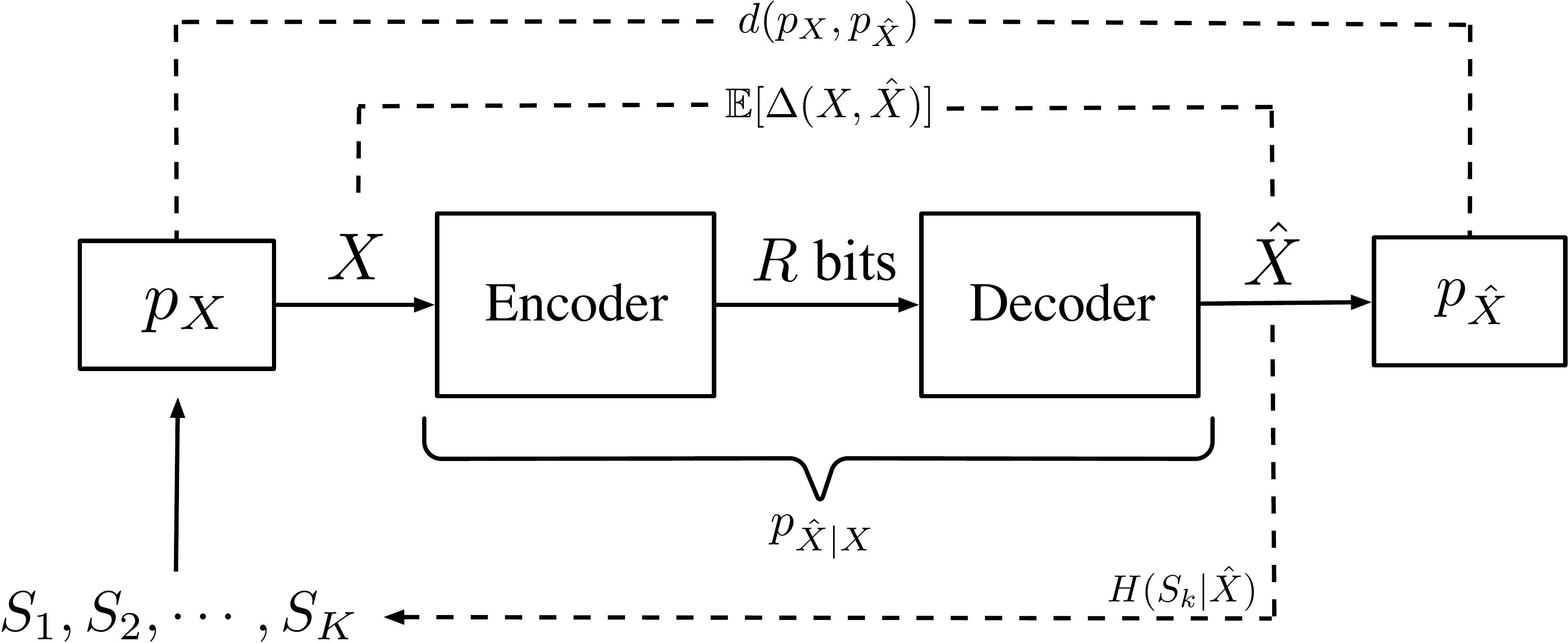}
  \caption{Illustration of task-oriented lossy compression framework.
  }
  \label{Fig_framework}
\end{figure}

 As shown in Fig. \ref{Fig_framework}, the process of lossy compression consists of an encoder and a decoder. 
 \begin{itemize}
  \item The encoding function maps the source $X$ to a message with a rate of $R$ bits. 
  \item The decoding function reproduces data $\hat X$ to satisfy task-oriented demands of downstream applications.
 \end{itemize}

 In task-oriented lossy compression, the destination may involve various tasks upon receiving a compressed signal, including traditional data reconstruction, generative learning tasks, or the prediction of classification labels. To accommodate these potential applications, we present the following symbol-level constraints. Theoretical analysis based on these constraints could reveal the practical usefulness of the generalized IB as an optimization objective\cite{IBappML}.
  
\emph{1) Reconstruction constraint:} We consider the following reconstruction constraint 
\begin{align*}
  \mathbb{E}(\Delta(X,\hat{X})) \leq D,
\end{align*}
where $\Delta:\mathcal X\times \hat{\mathcal X}\to \mathbb R^+$ is a data distortion function, such as Hamming distortion and squared-error distortion, and the expectation is taken over $p_{X, \hat X}(x, \hat x)=p_{\hat X|X}(\hat{x}|x)p_X(x)$.

\emph{2) Perception constraint:} The perceptual quality usually refer to the degree to  which an image resembles a natural image rather than a synthetic or restored image generated by an algorithm\cite{moorthy2011blind_perceptual}. It has been demonstrated that perceptual quality could be associated with the distance between the distributions of natural images and generated images \cite{PD-tradeoff}. Meanwhile, the underlying principle of generative adversarial network (GAN)\cite{goodfellow2014GAN} and its variants (e.g., Wasserstein GAN\cite{WGAN}) involve minimizing a divergence between two distributions. Hence, in this paper we adopt the same perception constraint as \cite{PD-tradeoff,RethinkingRDP,UniversalRDPs,CDP}
\begin{align*}
  d(p_X,p_{\hat{X}}) \leq P,
\end{align*}
where $d(\cdot, \cdot)$ is some divergence between probability distributions, such as total-variation (TV) divergence and Kulback-Leibler (KL) divergence.

\emph{3) Classification constraint:}  The conditional entropy $H(X|Y)$ measures the uncertainty in $X$ given information $Y$. Here we adopt the following classification constraint
\begin{align*}
  H(S_k|\hat{X}) \leq C_k, \qquad k\in[K],
\end{align*}
for some $C_k>0$, which means the uncertainty of classification variable  $S_k$ given the recovered source $\hat{X}$ should not surpass the level $C_k$. Equivalently, the constraints can be written as 
  $I(S_k,\hat{X}) \geq C^{'}_k,\forall k\in [K]$,
which indicates that a certain amount of semantic information about the relevant variable $S_k$ is preserved in $\hat X$. In Section \ref{SecExp_DLframework}, we will show that conditional entropy serves as a lower bound for the cross-entropy loss, a commonly employed metric in machine learning for optimizing classification models.

To characterize the achievable rate under all distortion, perception, and classification constraints, we can define the information rate-distortion-perception-classification (RDPC) function for a source $X\sim p_X(x)$ as
\begin{align}
    R(D,P,\mathbf{C}) = &\min_{p_{\hat{X}|X}} I(X;\hat{X})\label{RDPC}\\
    \quad\text{s.t.}~ 
    &\mathbb{E}[\Delta(X,\hat{X})] \leq D,\tag{\ref{RDPC}{a}} \label{Distortion}\\
    &d(p_X,p_{\hat{X}}) \leq P,\tag{\ref{RDPC}{b}} \label{Perception}\\
    &H(S_k|\hat{X}) \leq C_k, \forall k\in[K],\tag{\ref{RDPC}{c}} \label{Classification}
\end{align}
where $\mathbf{C}=(C_1,C_2,\cdots,C_K)$ is the allowed uncertainty of classification variables given $\hat X$ in classification constraints.

\emph{Connection with previous work:} The proposed $R(D,P,\mathbf{C})$  function can be reduced to a series of previous theories by relaxing one or two constraints: When the classification constraint \eqref{Classification} is relaxed, $R(D,P,\mathbf{C})$ is equivalent to the RDP function in \cite{RethinkingRDP}. If the perception constraint is also relaxed, the RDP function is reduced to the rate-distortion function $R(D)$ \cite{cover2012elements,yeung2008ITandNC}. When there exists only a single classification variable and both reconstruction  \eqref{Distortion} and perception constraints \eqref{Perception} are inactive, the proposed $R(D,P,\mathbf{C})$ is equivalent to information bottleneck principle\cite{IB}. Specifically, the problem is reduced to
  $\min_{p_{\hat{X}|X}:I(S,\hat{X}) \geq C} I(X;\hat{X})$.
By introducing the Lagrange multiplier $\beta$, the above problem is equivalent to
  $\min_{p_{\hat X|X}} \{I(X;\hat{X}) - \beta I(S,\hat{X})\}$,
which is the information bottleneck problem\cite{IB}.

\begin{remark}[Optimality with strong asymptotical constraints]\label{Remark_Operational}
    In the above formulation, we directly use the single letter representation and propose the information RDPC function to minimize $I(X;\hat{X})$ with diverse constraints. For multiple letters with i.i.d. sequence $X_1, X_2, \cdots, X_n\sim p(x)$, the operational meaning can be obtained from the results in \cite{CodingTheorem_RDP_Theis2021} via the Poisson functional representation\cite{Poisson_Functional} and common randomness exists between the encoder and decoder. The proof of both converse and achievability follows from demonstrating that the constraints in \eqref{Distortion}-\eqref{Classification} are functions of the joint distribution $p_{\hat X, X}$, thereby making the information RDPC function a specific case of the information rate function defined in \cite{CodingTheorem_RDP_Theis2021}. Note that asymptotical achievability in \cite{CodingTheorem_RDP_Theis2021} is very strong in the sense that all constraints should be satisfied by each single message.
\end{remark}
\begin{remark}[Achievability with weak asymptotical constraints]
  In a weak definition of achievability where the constraints are satisfied averagely (e.g., $\sum_{i}^N \mathbb{E}[\Delta(X_n,\hat{X}_n)]\leq D$ for some i.i.d process $\{X_n\}_1^N$ with margin $p_X$), the RDPC function is still achievable since the strong achievability implies the weak achievability.  
  Meanwhile, we can establish the optimality of RDC function by proving its convexity in general. For the RPC function, as demonstrated in Section \ref{sec-RPC}, it reduces to the RC case for the Bernoulli and scalar Gaussian sources, which are also convex.
\end{remark}
Nevertheless, this paper does not primarily focus on the operational definition of the RDPC function. Instead, its main objective is to provide insights into the tradeoffs between the rate and different tasks, and offer guidance for balancing diverse objectives in DL algorithms through the design of loss functions. The effectiveness of the RDPC function will be demonstrated with our experimental results in Section \ref{SecExperimentResults}.

\section{Investigating RDC and RPC Tradeoffs} \label{SecRDC-RPC}
In this section, we will initially examine two unexplored relationships by relaxing one constraint of the RDPC function: the rate-distortion-classification (RDC) and the rate-perception-classification (RPC).

\subsection{Rate-distortion-classification Tradeoff}\label{sec-RDC}
When relaxing perception constraint \eqref{Perception} and considering only one classification variable, we obtain the following information rate-distortion-classification function:
\begin{align}
  R(D,C) = &\min_{p_{\hat{X}|X}} I(X;\hat{X})\label{RDC}\\
    \quad\text{s.t.}~ 
    &\mathbb{E}(\Delta(X,\hat{X})) \leq D,\tag{\ref{RDC}{a}}\label{RDC-D}\\
    &H(S|\hat{X}) \leq C,\tag{\ref{RDC}{b}}\label{RDC-C}
\end{align}
where $S$ is a classification variable.

\subsubsection{Binary source}\label{sec-RDC-Bin}
For binary sources and classification labels,  we characterize the closed-form solution for $R(D,C)$ as the following theorem.
\begin{theorem}\label{TheoremBinaryRDC}
  Consider a Bernoulli source $X$ and a classification variable $S$ with the binary symmetric joint distribution given by $S=X\oplus S_1$ where $S\sim \text{Bern}(a)$ and $S_1\sim \text{Bern}(p_1)$ ($a,p_1\leq\frac{1}{2}$). The problem is infeasible if $C<H(S_1)$. Otherwise, the information rate-distortion-classification function with Hamming distortion is given by
  \begin{align*}
    R(D,C)= \begin{cases}
      H(b) - H(C_1) \qquad &\text{ for } D\geq C_1 \text{ and } C_1\leq b,\\
      H(b) - H(D) \qquad &\text{ for } D< C_1 \text{ and } D\leq b,\\
      0 \qquad &\text{ for } \min\{D,C_1\}> b,
    \end{cases}
  \end{align*}
  where $b=\min\{\frac{a-p_1}{1-2p_1}, 1-\frac{a-p_1}{1-2p_1}\}$ and $C_1= \frac{H^{-1}(C)-p_1}{1-2p_1}$. Here $H^{-1}:[0,1]\rightarrow [0,\frac{1}{2}]$ denotes the inverse function of Shannon entropy for probability less than $\frac{1}{2}$.
\end{theorem}

\begin{IEEEproof}
The main idea of this proof is to first utilize the rate-distortion theory\cite{cover2012elements} and Mrs. Gerber's Lemma \cite{MrsGerbers} to find a lower bound on the optimal rate. Then we prove that this lower bound is achievable. 
Please refer to Appendix \ref{Appendix_Proof_RDC_Bin} for more details.
\end{IEEEproof}

From the theorem, we can observe that when the distortion constraint $D$ is relatively large, the rate $R$ becomes a function of the classification constraint $C$, indicating that the classification constraint becomes the primary limiting factor. Conversely, when $D$ is relatively small and distortion becomes the dominant constraint, the rate $R$ becomes a function of the distortion constraint $D$.

In Fig. \ref{R-DC-Theoretical}(a), we visualize the result of Theorem \ref{TheoremBinaryRDC} by plotting the $R(D,C)$ function of a Bernoulli source as a surface (Fig. \ref{R-DC-Theoretical}(a), left) as well as distortion-classification curves for different rates (Fig. \ref{R-DC-Theoretical}(a), right). The first observation from the figures is that $R(D,C)$ function is non-increasing and convex over $D$ and $C$. When $D$ is relatively large, the value of rate is primarily determined by the value of $C$, and vice versa. When rate bits become larger, the overall curve shifts to the lower left, which means we can achieve a better classification and a better distortion level simultaneously.

\subsubsection{Gaussian source}\label{sec-RDC-Gaussian}
For the scalar Gaussian source,  we characterize the closed-form solution for $R(D,C)$ as the following theorem.
\begin{theorem}\label{TheoremRDCGS}
  Consider a Gaussian source $X\sim \mathcal{N}(\mu_x,\sigma_x^2)$ and a classification variable $S\sim \mathcal{N}(\mu_s,\sigma_s^2)$ with covariance $\text{Cov}(X,S)=\theta_1$. The RDC problem is infeasible if $C < \frac{1}{2}\log(1-\frac{\theta_1^2}{\sigma_s^2\sigma_x^2})+h(S)$. Otherwise, the information rate-distortion-classification function with MSE distortion is given by
  \begin{align*}
    R(D,C)= \begin{cases}
      \frac{1}{2}\log\frac{\sigma_x^2}{D} &\text{ for } D\leq \sigma_x^2(1-\frac{1}{\rho^2}(1-e^{-2h(S)+2C})),\\
      -\frac{1}{2}\log(1-\frac{1}{\rho^2}(1-e^{-2h(S)+2C}))& \text{ for } D> \sigma_x^2(1-\frac{1}{\rho^2}(1-e^{-2h(S)+2C})),\\
      0& \text{ for } C>h(S)\quad \text{and}\quad D>\sigma_x^2,
    \end{cases}
  \end{align*}
  where $\rho = \frac{\theta_1}{\sigma_s\sigma_x}$ is the correlation factor of $X$ and $S$, and $h(\cdot)$ is the differential entropy for a continuous variable.
\end{theorem}

\begin{IEEEproof}
  The converse and achievability proof is provided in Appendix \ref{Appendix_Proof_RDC_GS}.
\end{IEEEproof}

\begin{figure}[t]
  \centering
    \includegraphics[width=0.66\textwidth]{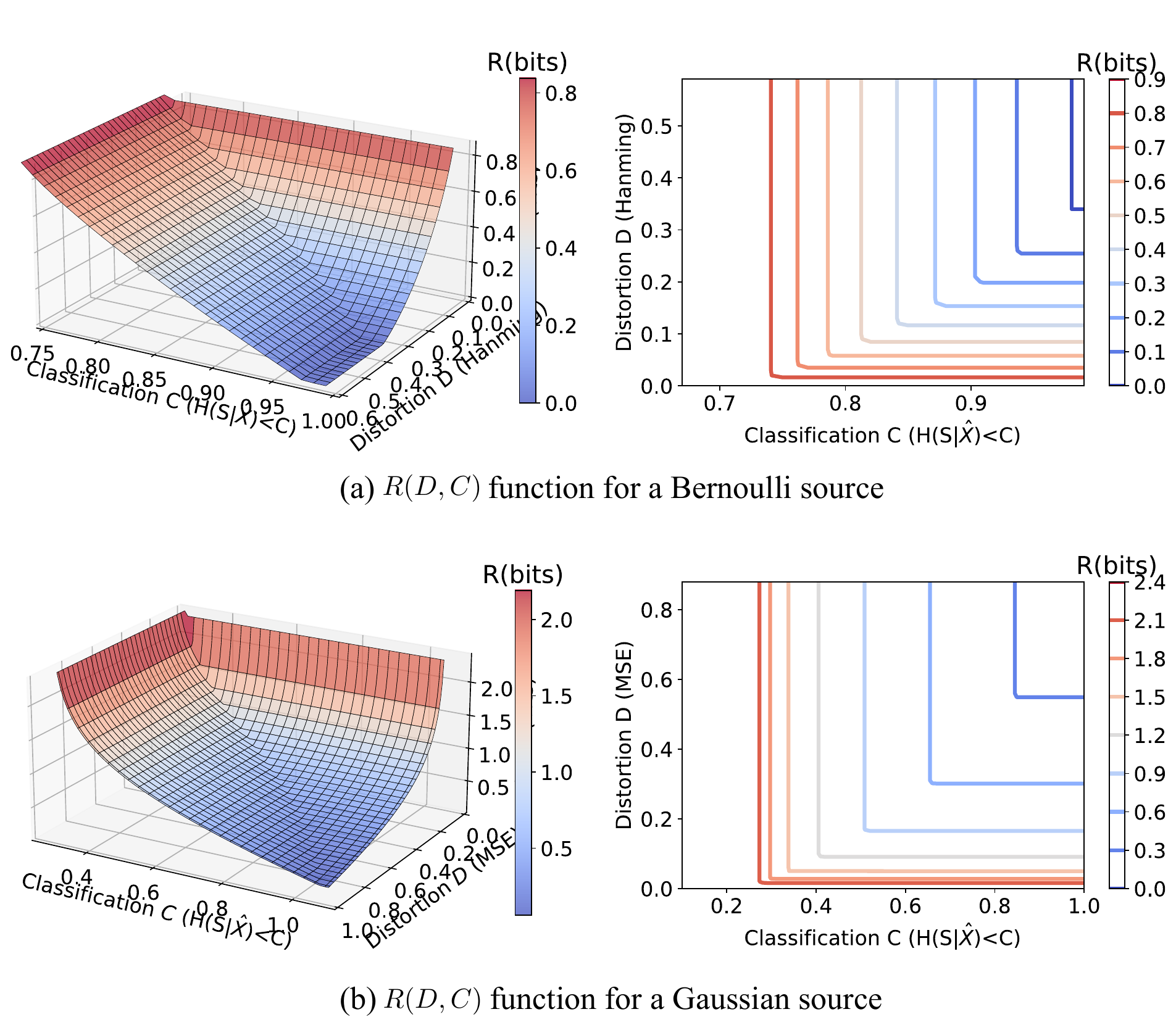}
  \caption{Visualization for RDC function of (a) a Bernoulli source and (b) a Gaussian source. The left figures depict $R(D,C)$ function as a surface. The colored solid lines on the figure indicate the different rate levels. The right figures show $R(D,C)$ function along distortion-classification planes, which is the two-dimensional projection of contour lines of the left figures.
  }
  \label{R-DC-Theoretical}
\end{figure}

We also visualize the result of Theorem \ref{TheoremRDCGS} by plotting the $R(D,C)$ function of a Gaussian source as a surface (Fig. \ref{R-DC-Theoretical}(b), left) as well as distortion-classification curves for different rates (Fig. \ref{R-DC-Theoretical}(b), right). Similar properties with binary case such as convexity and non-increasing of $R(D, C)$ on $D$ and $C$ are also observed.

\begin{figure}[!htbp]
  \centering
    \label{R-DC-RegionGS}
    \includegraphics[width=0.46\textwidth]{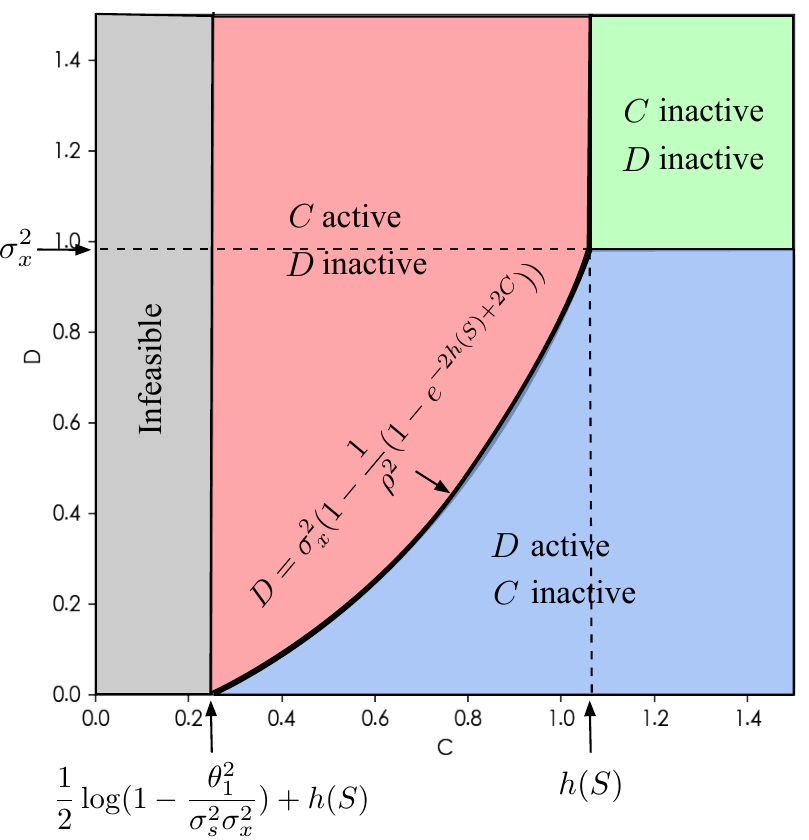}
  \caption{Visualization of the regions showing activeness of the distortion and classification constraints given each pair of $(D,C)$.
  }
  \label{R-DCregion}
\end{figure}

In addition, we investigate the activeness of distortion constraint \eqref{RDC-D} and classification constraint \eqref{RDC-C} given different value pairs of $(D,C)$.
As depicted in Fig. \ref{R-DCregion}, there is an 'antagonistic' relationship between D and C. When D is relatively small ($D\leq \sigma_x^2(1-\frac{1}{\rho^2}(1-e^{-2h(S)+2C}))$) and the distortion constraint is active, the classification constraint becomes inactive. Consequently, D predominates in determining the rate.
Conversely, as $D$ increases, the classification will be the only active constraint and the rate is determined by $C$. 

The phenomenon differs from the case of RDP \cite{RethinkingRDP}\cite{UniversalRDPs}\cite{RDP-OP2024}. As proved in \cite{UniversalRDPs}, in the Gaussian case, when $\sqrt{P}\geq \sigma_X-\sqrt{|\sigma_x^2-D|}$, the distortion constraint becomes the only active constraint, and the rate is solely determined by $D$. However, as $P$ decreases, both $P$ and $D$ are active, leading to a joint determination of the rate by $D$ and $P$. Here $D$ is always active since $I(X,\hat X) = 0$ is achievable for any $P$ when $D$ is not an active constraint\cite{RethinkingRDP}.

\subsection{Rate-perception-classification Relationship}\label{sec-RPC}

Unlike the cases of RDP and RDC, we will show that there is no tradeoff between perception and classification in the RPC relationship for both binary and Gaussian scenarios.   Intuitively, the restriction on target label inference may not affect the potential of perfect perceptual quality. 

Specifically, let us consider the following information rate-perception-classification function where the distortion constraint \eqref{Distortion} is relaxed in the RDPC function and only one classification variable is considered:
\begin{align}
  R(P,C) = &\min_{p_{\hat{X}|X}} I(X;\hat{X})\label{RPC}\\
    \quad\text{s.t.}~ 
    &d(p_X,p_{\hat{X}}) \leq P,\tag{\ref{RPC}{a}}\label{RPC-P}\\
    &H(S|\hat{X}) \leq C,\tag{\ref{RPC}{b}}\label{RPC-C}
\end{align}
where $S$ is the only classification variable.

\subsubsection{Binary source}\label{sec-RPC-Bin}

The following theorem characterizes $R(P,C)$ for a binary source with perception quality described by total variation divergence.
\begin{theorem}\label{TheoremBinaryRPC}
  Consider a Bernoulli source $X$ and a classification variable $S$ with the binary symmetric joint distribution given by $S=X\oplus S_1$ where $S\sim \text{Bern}(a)$ and $S_1\sim \text{Bern}(p_1)$ ($a,p_1\leq 1/2$). If $C<H(S_1)$, the problem is infeasible. Otherwise, the information rate-perception-classification function with total variation (TV) divergence is given by
  \begin{align*}
    R(P,C)= \begin{cases}
      H(b) - H(C_1) \qquad &\text{ for } H(S_1)\leq C\leq H(S),\\
      0 \qquad &\text{ for } C> H(S),
    \end{cases}
  \end{align*}
  where $b=\min\{\frac{a-p_1}{1-2p_1}, 1-\frac{a-p_1}{1-2p_1}\}$ and $C_1= \frac{H^{-1}(C)-p_1}{1-2p_1}$.
\end{theorem}

\begin{IEEEproof}
  The conditional entropy constraint provides a lower bound on the rate. By delicately assigning the value of conditional probability $P(\hat X=0|X=0)$ and $P(\hat X=0|X=1)$, the lower bound is achievable with the TV divergence $d_{TV}(p_X,p_{\hat X})=0$. See details in Appendix \ref{Appendix_Proof_RPC_Bin}.
\end{IEEEproof}

The above theorem indicates that there is no tradeoff between perception and classification for the binary case, since the rate only depends on the classification constraint. The proof of Theorem \ref{TheoremBinaryRPC} establishes that it is always possible to find an optimal solution that achieves a TV divergence of zero while attaining the lower bound of the rate imposed by the classification constraint.

\subsubsection{Gaussian source}\label{sec-RPC-Gaussian}
The conclusions in the binary case can be extended to the Gaussian case, where the optimal rate is solely determined by the level of the conditional entropy constraint, and the perception constraint is always inactive. In this subsection, we take the natural base of logarithm.

\begin{theorem}\label{TheoremRPCGS}
  Consider a scalar Gaussian source $X\sim \mathcal{N}(\mu_x,\sigma_x^2)$ and a classification variable $S\sim \mathcal{N}(\mu_s,\sigma_s^2)$ with covariance $\text{Cov}(X,S)=\theta_1$. If $C < \frac{1}{2}\log(1-\frac{\theta_1^2}{\sigma_s^2\sigma_x^2})+h(S)$, the problem will be infeasible. Otherwise, the rate-perception-classification function with KL divergence is given by
  \begin{align*}
    R(P,C)= \begin{cases}
      -\frac{1}{2}\log(1-\frac{1}{\rho^2}(1-e^{-2h(S)+2C})),& \text{ for } \frac{1}{2}\log(1\!-\!\rho^2)\!+\!h(S)\leq C\leq h(S);\\
      0,&\text{ for } C>h(S);
    \end{cases}
  \end{align*}
  where $\rho = \frac{\theta_1}{\sigma_s\sigma_x}$ is the correlation factor of $X$ and $S$, and $h(\cdot)$ is the differential entropy for a continuous variable.
\end{theorem}

\begin{IEEEproof}
  In Appendix \ref{Appendix_Proof_RPC_GS}, we will prove that the lower bound given by the conditional entropy constraint is achievable with the KL divergence always zero.
\end{IEEEproof}

Intuitively, without distortion constraint, we can generate arbitrary reconstructions with the same distribution as the source (even independent of the original data) to obtain perfect perception. Conditional entropy constraint requires preserving the covariance of $S$ and $X$, which intuitively does not contradict preserving the distribution of $X$. Although our theoretical results focus on Binary and Gaussian cases, similar phenomena will be observed in experiments of real-world datasets in Section \ref{SecExperimentResults}.

\begin{remark}
  An interesting observation is that RDC, RPC, and the previous RDP\cite{RethinkingRDP} exhibit distinct characteristics within each trinary relationship for both binary and Gaussian case. Specifically, for the RDP case, the distortion constraint is always active as long as $R(D,P)>0$\cite{RethinkingRDP}, while the activeness of perception depends on its relative value with distortion. For the RDC case, under a limited rate, distortion and classification are at odds with each other in an `antagonistic' way where at most one constraint could be active at the same time. Finally, when considering the perception and classification constraints, the optimal rate solely depends on the classification and is irrelevant to the perception. 
\end{remark}

  \begin{remark}
    The results derived from Theorems \ref{TheoremBinaryRDC}-\ref{TheoremRPCGS} can be naturally extended to $K$ classification variables if $S_1,\cdots, S_K$ are independent. In the binary case, imposing multiple constraints $H(\hat{S}_k|\hat{X})\leq C_k$ for $k=1,2,\cdots, K$ results in different bounds on $H(X|\hat{X})\leq H(\tilde{C}_k)$, where $\tilde{C}_k=\frac{H^{-1}(C_k)-p_1}{1-2p_1}$ according to Mrs. Gerber's Lemma. It is equivalent to take the most stringent constraint $\tilde{C}\triangleq \min\{\tilde{C}_1,\cdots,\tilde{C}_k\}$ and determine the lower bound of $I(X,\hat{X})$. Then, employing analogous techniques as in the proofs of Theorems \ref{TheoremBinaryRDC} and \ref{TheoremBinaryRPC} yields similar outcomes as those for a single label. Likewise, we can also establish the equivalence between multiple constraints and the tightest constraint in the Gaussian case, so that Theorems \ref{TheoremRDCGS} and \ref{TheoremBinaryRPC} can be extended to multiple labels. Exploring scenarios involving correlated classification variables is regarded as a future research direction.
\end{remark}

  The proposed RDC and RPC functions complement the IB principle and RDP tradeoff, providing new insights regarding the interplay between different tasks under a limited rate. Specifically, the perceptual quality primarily relates to the distribution of the data, which does not inherently create a tradeoff with the rate. 
However, when constrained by a limited rate, \emph{the distortion and classification constraints have distinct impacts on the perceptual quality}. That is, minimizing distortion potentially leads to degradation of perceptual quality, while imposing a conditional entropy constraint does not limit the ability to perfectly preserve the distribution, at least in the binary and Gaussian scenarios. 

On the other hand, when the rate is not limited, it becomes possible to reconstruct the source information perfectly. So, there should be no tradeoffs between the different tasks if the rate is sufficiently large. However, within the framework of signal restoration, previous literature\cite{CDP} has demonstrated the existence of a classification-distortion-perception (CDP) tradeoff without an explicit constraint on the communication rate. In the next section, we will initiate a discussion to identify the critical factor that determines the presence of tradeoffs.

\section{The Decisive Role of Source Noise in Tradeoffs} \label{SecDistortion}
In this section, we will first compare the models of lossy compression and signal restoration. For the latter, an extrinsic noise is first introduced to degrade the source $X$ and the goal is to restore $\hat X$ from the degraded observation $Y$. With a toy example, we will illustrate that in the scenario of \emph{non-zero} noise level, the tradeoffs between distortion, perception and classification emerge. Then, we show that introducing an extrinsic noise is equivalent to imposing a certain level of distortion between $X$ and $\hat X$. With this equivalence, we further consider the RPC relationship given a specific level of distortion and observe different characteristics compared to the original RPC tradeoff.  The practical role of these two tradeoffs will be verified by the experiments in Section \ref{SecExperimentResults}. 

\subsection{Comparison of Lossy Compression and Signal Restoration}\label{sec-comparison_lossy_degra}

\begin{figure}[!htbp]
  \centering
    \includegraphics[width=0.45\textwidth]{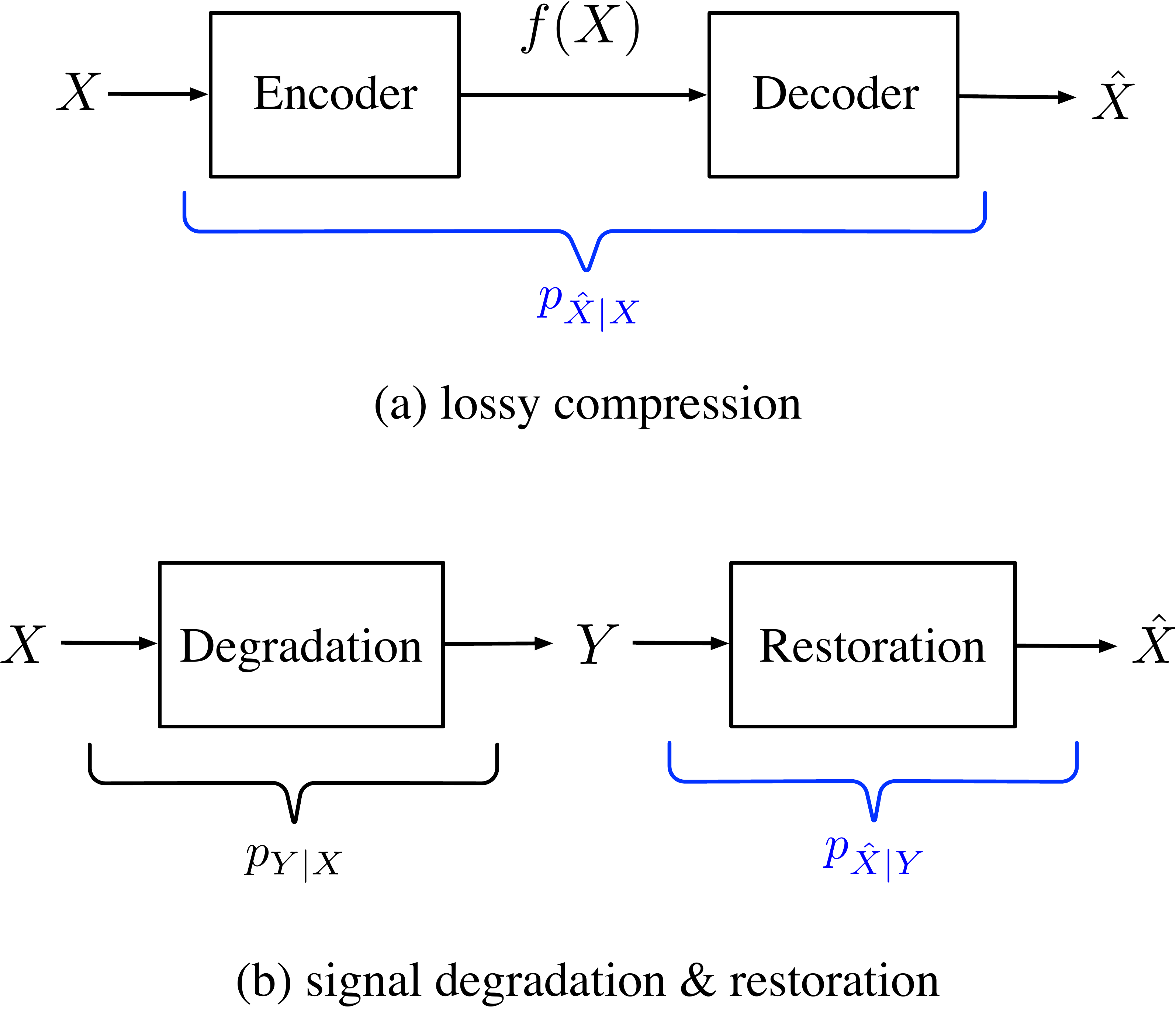}
  \caption{Frameworks of lossy compression model and signal restoration model.
  }
  \label{LC-and-SD}
\end{figure}

Fig. \ref{LC-and-SD} illustrates a comparison between the models of the lossy compression and the signal restoration. In the lossy compression setting, our objective is to find an efficient representation $f(X)$ that minimizes the communication bits given a constraint of distortion (or perception/classification in the task-oriented model). From an information-theoretic perspective, we directly optimize the conditional distribution $p_{\hat X|X}$ by jointly designing the encoder and decoder. In the signal degradation and restoration model, an extrinsic noise is first introduced through $p_{Y|X}$ to the source $X$, resulting in the observation of a degraded source $Y$. The task is to reconstruct $\hat X$ from the degraded $Y$. In the following, we mainly consider the additive Gaussian noise, i.e., $Y=X+N$ with $N\sim\mathcal N(0,\sigma_N^2)$. To delve deeper into the tradeoffs in the signal restoration model, let us begin by considering the toy example in \cite{CDP}.

\begin{example}[Toy example in \cite{CDP}]\label{Toy_example}
  Consider signal $X$ with Gaussian mixture distribution $p_X(x) = P_1p_{X_1}(x)+P_2p_{X_2}(x)$, where $P_1=0.7, P_2=0.3$, $p_{X_1}(x)=\mathcal N(-1,1)$, $p_{X_2}(x)=\mathcal N(1,1)$. The signal degradation is given by $Y=X+N$ where $N\sim\mathcal{N}(0,\sigma_N^2)$ with $\sigma_N=1$. Consider the linear denoising method $\hat{X}=aY$ where $a$ is an adjustable parameter. 

  According to the Bayes decision rule, the classification error rate given by the optimal classification plane $c_0$ is
  \begin{align*}
    \varepsilon(\hat{X}|c_0) &= P_2\int_{-\infty}^{x_0}p_{\hat X_2}(x)dx +  P_1\int_{x_0}^{\infty}p_{\hat X_1}(x)dx  \\
    &= P_2\Phi(x_0^\prime) + P_1\Phi(-x_0^{\prime\prime}),
  \end{align*}
  where $x_0^\prime=\frac{x_0-a}{|a|\sqrt{1+\sigma_N}}$ and $x_0^{\prime\prime}=\frac{x_0+a}{|a|\sqrt{1+\sigma_N}}$, and $\Phi(\cdot)$ is the cumulative distribution function of the standard normal distribution. Note that the error rate is a function of $a$.

  Meanwhile, the KL divergence $d_{KL}(p_X,p_{\hat X})$ and MSE $\mathbb E[\Delta(X,\hat X)]$ are also functions of the denoising parameter $a$ which can be computed numerically. 
  
  Specifically, MSE as a function of $a$ is plotted in Fig. \ref{top_example_CDP}.
  Note that the optimal MSE is $0.204$ obtained at $a=0.67$. No matter how we choose the value of $a$, MSE will never be 0, which means perfect reconstruction is impossible.

  \begin{figure}[!htbp]
    \centering
      \includegraphics[width=0.35\textwidth]{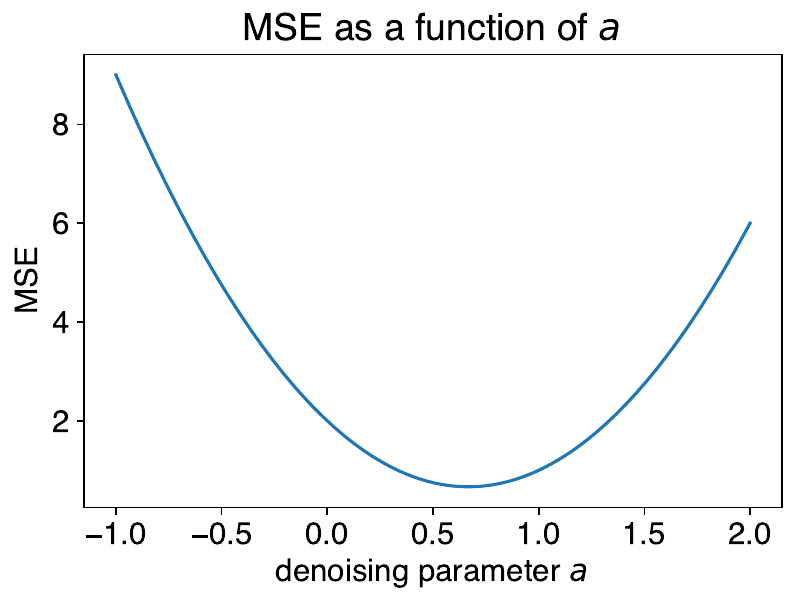}
    \caption{MSE as functions of linear denoising parameter $a$ in the toy example.
    }
    \label{top_example_CDP}
  \end{figure}

  Meanwhile, all three relationships, namely distortion-perception (DP), distortion-classification (DC), and perception-classification (PC) exhibit tradeoffs, as shown in Fig. \ref{top_example_tradeoffs}. For example, in the case of the DP tradeoff shown in Fig. \ref{top_example_tradeoffs}(a), we draw the pairs $(D,P)$ as the solution of
  \begin{align*}
    P=&\min_a \ d_{KL}(p_X,p_{\hat X})\\
    &\text{s.t.} \quad \mathbb E[(x-\hat x)^2] \leq D,
  \end{align*}
  and it shows that the decrease in MSE is accompanied by an increase in KL divergence.

  \begin{figure}[!htbp]
    \centering
      \includegraphics[width=0.56\textwidth]{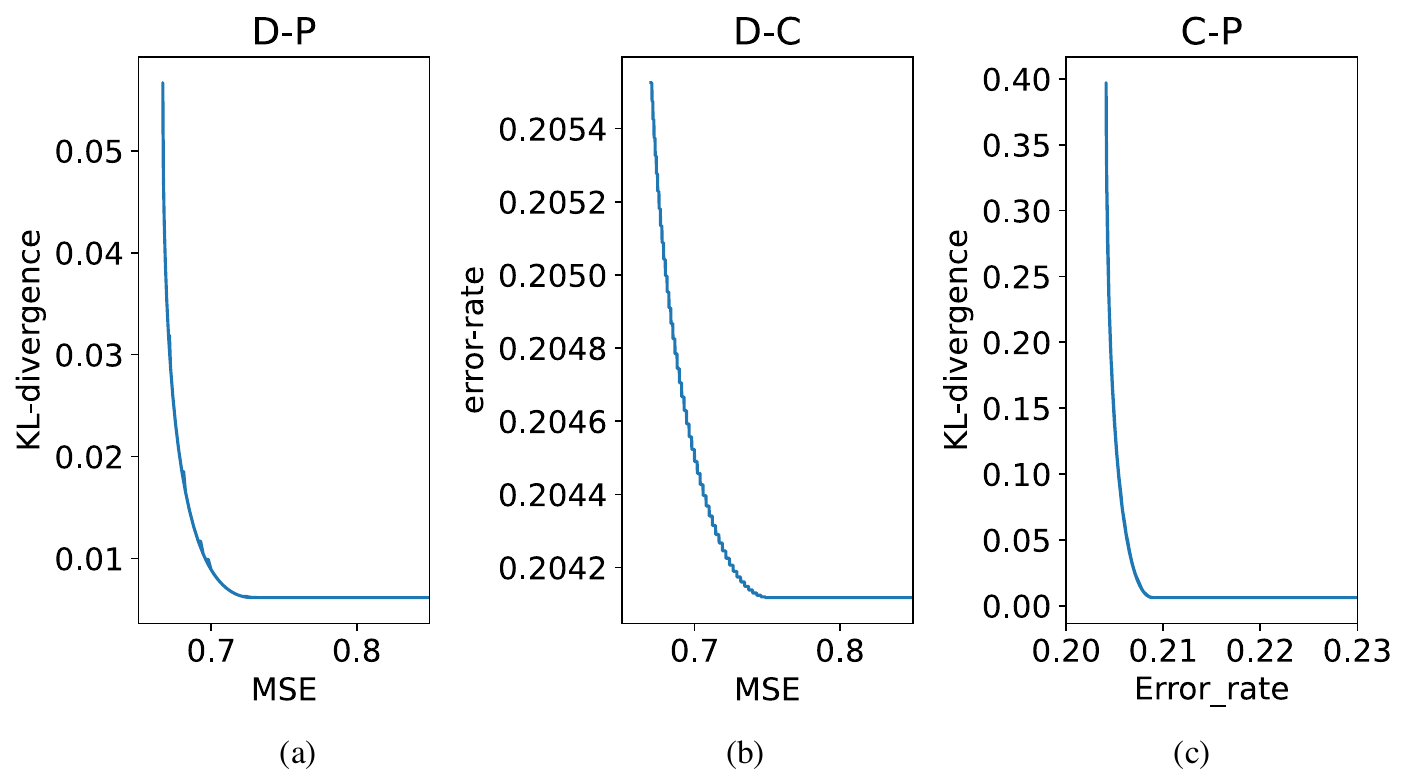}
    \caption{Distortion-perception tradeoff (a), distortion-classification tradeoff (b), and perception-classification tradeoff (c) in the toy example.}
    \label{top_example_tradeoffs}
  \end{figure}
\end{example}

\begin{figure}[!htbp]
  \centering
    \includegraphics[width=0.68\textwidth]{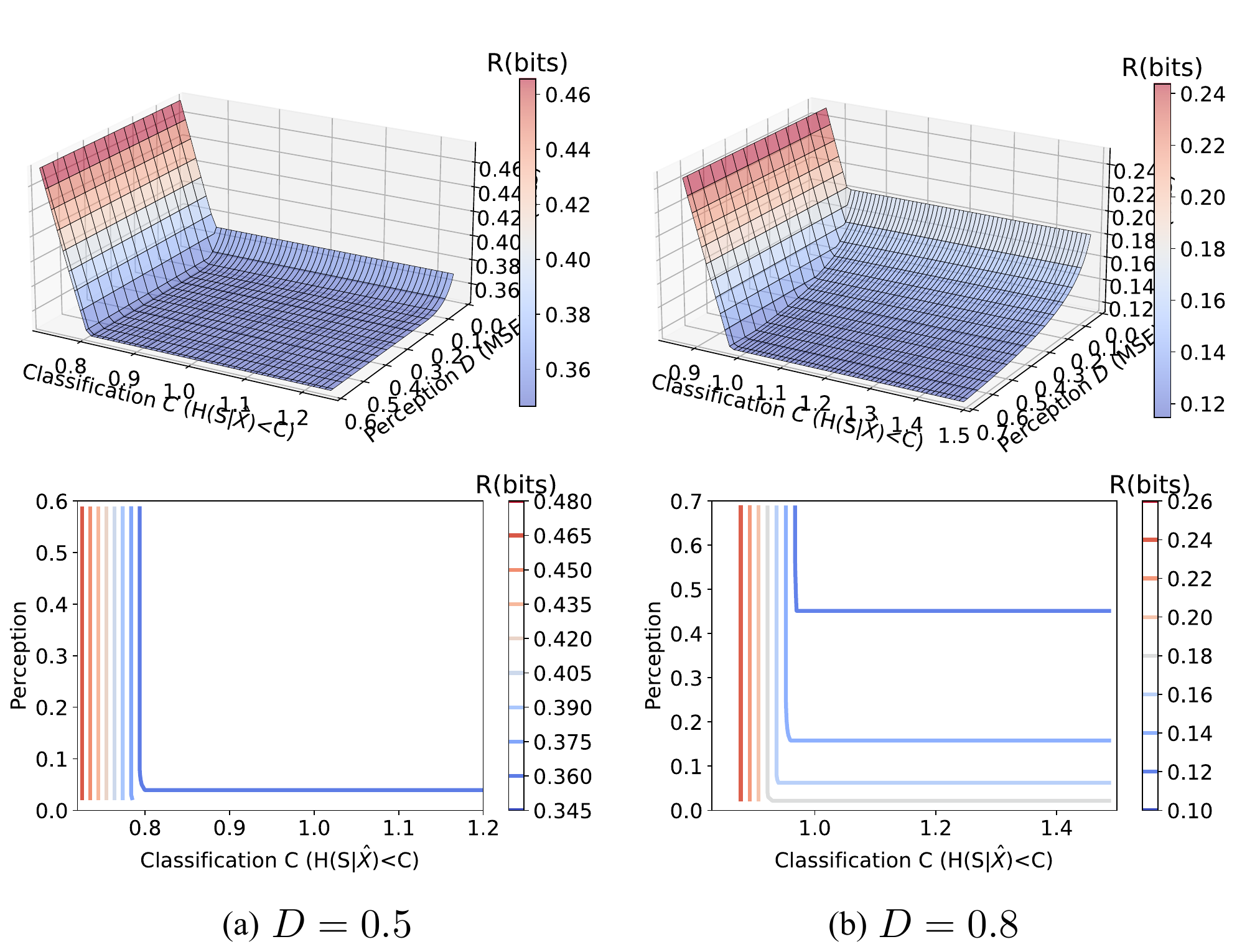}
  \caption{Simulation of rate-perception-classification function given a certain level of distortion $R(P,C|D)$ in the Gaussian case where $\sigma_x = 1, \sigma_s=0.7, \theta_1=0.63$. (a) The given distortion $D=0.5$; (b) The given distortion $D=0.6$; (c) The given distortion $D=0.8$.
  }
  \label{RPC-D-Plot}
\end{figure}

Similar to the case of the sufficiently large rate in the lossy compression framework, when the noise level $\sigma_N$ tends to zero here, there are no tradeoffs between different tasks. In this scenario, we directly observe the original data. The identity restoration by setting $a=1$ yield with zero MSE, zero KL divergence, and the optimal classification performance.

However, when dealing with a non-zero noise level, we observe that the objectives of minimizing MSE, error rate, and KL divergence do not always align, leading to tradeoffs between each pair of constraints. Specifically, for a linear denoiser, the optimal MSE is achieved at $a=0.67$, the minimum error rate is achieved at $a=0.50$, and the lowest KL-divergence is obtained at $a=0.81$ when $\sigma_N=1$. The value of optimal parameter $a$ depends on the noise level $\sigma_N$.

In the following, we show that introducing an non-zero extrinsic noise is equivalent to imposing a certain level of distortion between $X$ and $\hat X$. Starting with a simple example, where the degradation is given by $Y=X+N\sim\mathcal{N}(0,1+\sigma_N^2)$ and the restoration is given by $\hat X=aY\sim\mathcal{N}(0,a^2(1+\sigma_N^2))$, for distortion measured in MSE, we have
\begin{align*}
  \mathbb E[\Delta(X,\hat X)]&=1+a^2(1+\sigma_{N}^2)-2a=(1+\sigma_N^2)(a-\frac{1}{1+\sigma_N^2})^2+\frac{\sigma_N^2}{1+\sigma_N^2},
\end{align*}
which is always greater than zero as long as $\sigma_N\neq 0$.
In general, based on the results from lossy compression with a noisy source \cite{DobrushinTsybakov_noisy_information1962,WolfZiv_noisy_information1970}, when the distortion is measured by MSE, we have
\begin{align*}
  \mathbb E[||X\!-\!\hat X||^2] = \mathbb E[||X-\mathbb E[X|Y]||^2] + \mathbb E[||\hat X\!-\!\mathbb E[X|Y]||^2],
\end{align*}
where the term $\mathbb E[||X-\mathbb E[X|Y]||^2]$ determines the minimum distortion, which is independent of the encoder-decoder pair.

Thus, the presence of source noise introduces an implicit constraint on distortion, making perfect reconstruction impossible. Detailed investigation of the tradeoffs inherent in lossy compression with a noisy source remains a topic for future research. In the following, we verify the significant role of source noise in the existence of tradeoffs from the perspective of imposing equivalent distortion constraints. Specifically, we reexamine the RPC relationship in Section \ref{sec-RPC} with a specific level of distortion. Through this analysis, we observe that the tradeoff between perception and classification indeed exists.

\subsection{RPC Given a Certain Level of Distortion}\label{sec-RPC_D}

In section \ref{sec-RPC}, we found that there is no tradeoff between perception and classification in the RPC relationship.
The optimal rate only depends on the classification constraint.
However, our subsequent simulations in the scalar Gaussian case will demonstrate that imposing a constraint on the distortion level causes a tradeoff to emerge.

Without loss of optimality, the RPC function for a scalar Gaussian case is equivalent to (see Appendix \ref{Appendix_Proof_RPC_GS})
\begin{align*}
  R(P, C) = &\min_{\sigma_{\hat{x},}\theta_2} -\frac{1}{2}\log(1-\frac{\theta_2^2}{\sigma_x^2\sigma_{\hat{x}}^2})\\
    \quad\text{s.t.}~ 
    &\frac{1}{2}\log\frac{\sigma_{\hat{x}}^2}{\sigma_{x}^2}+\frac{\sigma_x^2-\sigma_{\hat{x}}^2}{2\sigma_{\hat{x}}^2} \leq P,\\
    &-\frac{1}{2}\log(1-\frac{\theta_1^2}{\sigma_s^2\sigma_x^4}\cdot\frac{\theta_2^2}{\sigma_{\hat{x}}^2}) \geq h(S) - C,
\end{align*}
where $\sigma^2_{\hat X}$, $\theta_1$ and $\theta_2$ denote the variance of $\hat X$, $Cov(S, X)$ and $Cov(\hat X, X)$ respectively. If we consider a specific level of $D$, i.e., $\mathbb E[||X-\hat X||^2]=\sigma_x^2+\sigma_{\hat{x}}^2-2\theta_2 = D$, by substituting $\theta_2 = \frac{\sigma_x^2+\sigma_{\hat{x}}^2-D}{2}$ to the RPC function, we have
\vspace{-0.3cm}
\begin{align*}
  R(P,C|D) = \min_{\sigma_{\hat{x}}} &-\frac{1}{2}\log\big(1-\frac{(\sigma_x^2+\sigma_{\hat{x}}^2-D)^2}{4\sigma_x^2\sigma_{\hat{x}}^2}\big)\\
  \text{s.t.}\  &\frac{1}{2}\log\frac{\sigma_{\hat{x}}^2}{\sigma_{x}^2}+\frac{\sigma_x^2-\sigma_{\hat{x}}^2}{2\sigma_{\hat{x}}^2} \leq P,\\
  & \!\!\!\!\!\!\!\!\!\!\!\!\!\!\!\!-\!\frac{1}{2}\log\big(1\!-\!\frac{\theta_1^2}{\sigma_s^2\sigma_x^4}\!\cdot\!\frac{(\sigma_x^2\!+\!\sigma_{\hat{x}}^2\!-\!D)^2}{4\sigma_{\hat{x}}^2}\big) \!\geq\! h(S) \!-\! C.
  \end{align*} 
  Since the objective and two constraints are all functions of $\sigma_{\hat x}$, we can depict the curve of $R(P,C|D)$ by simulation, as shown in Fig. \ref{RPC-D-Plot}. Here, we choose $\sigma_x=1, \sigma_s=0.7,\theta_1=0.63$ and $D=0.5,0.88$ respectively.

  It is observed that the perception and classification exhibit a tradeoff, which shrinks as $D$ decreases. This can be explained by considering the interplay of $D$ with classification and perception respectively.
  From the Gaussian RDP tradeoff\cite{UniversalRDPs}, we know that when $P> \sigma_x-\sqrt{|\sigma_x^2-D|}$, the perception constraint is inactive and the rate only depends on $D$. Meanwhile, in Section \ref{sec-RDC-Gaussian}, we have shown that when $C\geq \frac{1}{2}\log(1-\rho^2(1-\frac{D}{\sigma_x^2}))+h(S)$, the rate only depends on $D$. Hence, when $D\rightarrow 0$, a perfect reconstruction is expected, and the rate is dominated by $D$. When $D$ is a bit larger and a degraded reconstruction is expected, the objectives of optimizing perception and classification are not always aligned and the rate will depend on the activeness of $C$ or $P$ constraints. Thus, we can see that the tradeoff emerges.

\section{Experimental Results} \label{SecExperimentResults}
 
In this section, we validate our theoretical results by implementing a DL-based image compression framework to achieve multiple objectives related to distortion, perception, and classification. 
These experiments not only verify the effectiveness of our theoretical results but also provide insights into loss function design in multi-task learning.

\begin{figure}[!htbp]
  \center \includegraphics[width=0.66\textwidth]{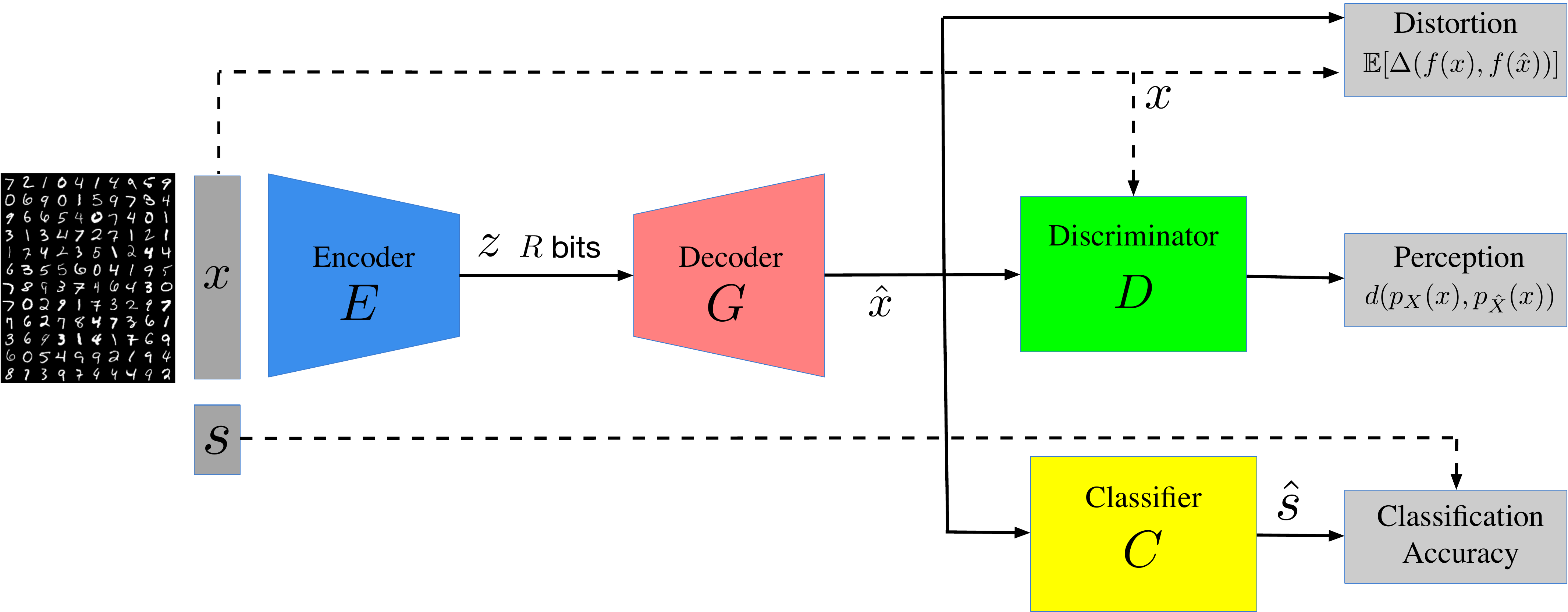}
    \caption{Illustration of rate-distortion-perception-classification network}\label{net}
    \vspace{-0.5cm}
\end{figure}

\subsection{DL-based Lossy Compression with RDPC Constraints} \label{SecExp_DLframework}
By utilizing a generative adversarial network (GAN), the process of lossy compression $p_{\hat{X}|X}(\hat{x}|x)$ can be achieved with a encoder and a decoder parameterized by adjustable parameter $\phi_e$ and $\phi_d$ respectively. Let $\phi=\{\phi_e,\phi_d\}$ denote the parameters of the whole encoder-decoder network.

Similar to \cite{UniversalRDPs}, we use a GAN-based network with stochastic encoder and decoder with universal quantization\cite{On_universal_quantization1985,Universally_quantized_neural_compression}. As shown in Fig. \ref{net}, the system consists of an encoder $E$, a decoder $G$, a discriminator $D$, and a pre-trained classifier $C$.

The reconstruction $\hat{x}$ and the original image $x$ are fed to the discriminator $G$ to compute the discriminator loss. The reconstruction $\hat{x}$ is also fed to a pre-trained classifier to obtain a vector $\hat{\mathbf{s}}$ with each entry being the probability that the image belongs to each class. The true label $s$ and the predicted $\mathbf{\hat{s}}$ are used to compute the classification accuracy.

To compute the conditional entropy constraints, we need to track the posterior distribution $p_{\phi}(s|\hat{x})$, which can be replaced by the pre-trained classifier parameterized by $\psi$. 
By introducing the approximation $p_{\psi}(s|\hat{x})$, we can derive an upper bound of the conditional entropy constraint\cite{boudiaf2021unifying_cross_entropy}:
\begin{align*}
    H(S|\hat{X})\! &= \sum_{s}\sum_{\hat{x}} p_{\phi}(s,\hat{x})\log\frac{1}{p_{\phi}(s|\hat{x})}\\
    &= \underbrace{ \sum_{s}\!\sum_{\hat{x}} \!p_{\phi}(s,\!\hat{x})\!\log\!\frac{p_{\psi}(s|\hat{x})}{p_{\phi}(s|\hat{x})} }_{ -d_{KL}(p_{\phi}(s|\hat{x})||p_{\psi}(s|\hat{x})) } \!+\!\! \underbrace{ \sum_{s}\!\sum_{\hat{x}} p_{\phi}(s,\hat{x})\!\log\!\frac{1}{p_{\psi}(s|\hat{x})} }_{\text{CE}(s,\mathbf{\hat{s}})}\\
    &\leq \text{CE}(s,\mathbf{\hat{s}}),
\end{align*}
where the inequality follows from the non-negativity of KL- divergence, and $\text{CE}(s,\mathbf{\hat{s}})$ is the \emph{conditional} cross-entropy loss. 

\begin{remark}
  In practical loss implementations, such as in PyTorch, people are using the same conditional cross-entropy loss as detailed above. Recall that $\hat{\mathbf{s}}$ is a vector output from the pre-trained classifier given $\hat{x}$, with the $i$ entry being the conditional probability of $i$th class, i.e., $p_{\psi}(s=i|\hat{x})$. Given parameters $\phi$, $\hat{x}=G_{\phi}(E_{\phi}(x))$ is a function of $x$. Denote the label of $x$ by $i_x$ and the batch size by $N$. Then $p_{\phi}(s,\hat{x})=p_{\phi}(s|G_{\phi}(E_{\phi}(x)))p_{\phi}(\hat{x})=1/N$ if $s=i_x$, and $p_{\phi}(s,\hat{x})=0$ otherwise. Thus, the cross-entropy loss can be expressed as 
\begin{align*}
  \text{CE}(s,\mathbf{\hat{s}})&=\sum_{s}\!\sum_{\hat{x}} p_{\phi}(s,\hat{x})\!\log\!\frac{1}{p_{\psi}(s|\hat{x})} = -\frac{1}{N}\sum_{\hat{x}}\log p_{\psi}(i_x|\hat x),
\end{align*}
which is the implementation used in PyTorch.
\end{remark}

For the reconstruction constraint, we use MSE $\mathbb{E}(||X-\hat{X}||^2)$ serving as the traditional distortion loss. For the perception constraint, we employ the Wasserstein-1 loss to measure the distance between distributions.

To control the compression rate $R$, we let $R$ be the upper bound $\text{dim}\times \log_2(L)$, where dim is the dimension of the encoder's output, and $L$ is the number of levels used for quantizing each entry. As discussed in \cite{RethinkingRDP}, setting $R$ to its upper bound significantly simplifies the scheme and is found to be only slightly sub-optimal \cite{GANImageCompress}.

In summary, when the rate $R=\text{dim}\times \log_2(L)$, the overall loss function of the DL-based lossy compression framework based on the RDPC function is
\begin{align}
    \mathcal L=\lambda_d \mathbb{E}(||X-\hat{X}||^2) + &\lambda_p W_1(p_X,p_{\hat{X}})+\lambda_c \text{CE}(s,\mathbf{\hat{s}}),\label{total-loss}
  \end{align}
where $\lambda_d, \lambda_p,$ and $\lambda_c$ are hyperparameters to control weights of distortion, perception, and cross-entropy losses, respectively.

\subsection{RDC and RPC Relationships}

\subsubsection{RDC tradeoff} To illustrate the relationship within rate-distortion-classification, we set the parameters in the loss function \eqref{total-loss} as $\lambda_p=0$ and train the model with a series of $\lambda_d$ and $\lambda_c$. For the MNIST dataset, rates are controlled by $R=\text{dim}\times\log_2(L)$ with (dim, $L$) pairs including $(3,3)$, $(3,4)$, $(4,4)$, $(5,4)$. As shown in Fig. \ref{RDC-fig}(a), each point represents a encoder-decoder pair trained with a specific combination of $R$, $\lambda_d$, and $\lambda_c$. For points with the same color (i.e., trained with the same $R$), the results show that higher classification accuracy often requires sacrificing distortion, which coincides with our theoretical results. It is also observed that $D-C$ curves shift to the left-up when $R$ increases, achieving better distortion and higher classification accuracy simultaneously.

\subsubsection{RPC relationship} 
\begin{figure}[tbp]
  \center \includegraphics[width=0.68\textwidth]{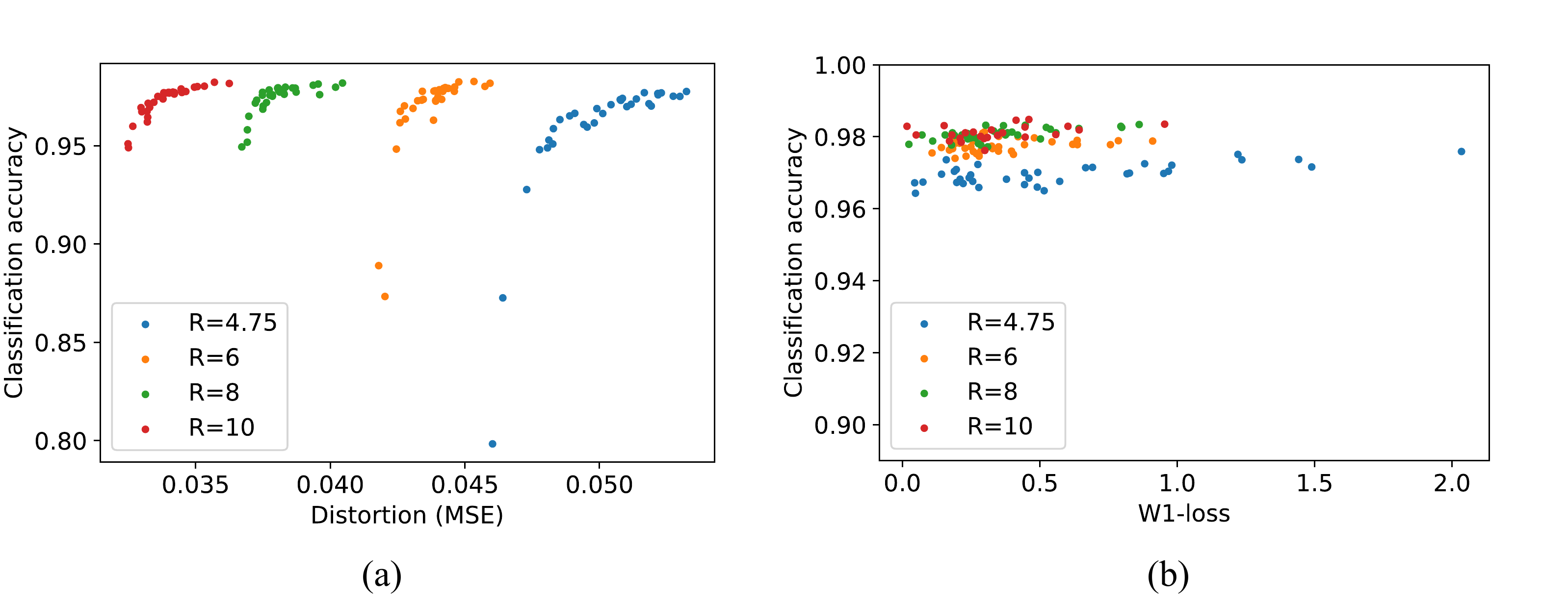}
    \caption{RDC and RPC tradeoffs in MNIST dataset. (a) For the RDC tradeoff, the loss is computed by setting $\lambda_p=0$ and varying $\lambda_d$ and $\lambda_c$. (b) For the RPC case, The loss is computed by setting $\lambda_d=0$ and varying $\lambda_p$ and $\lambda_c$.}\label{RDC-fig}
\end{figure}

By setting the parameters in the loss function \eqref{total-loss} as $\lambda_d=0$ and training the model with a series of $\lambda_p$ and $\lambda_c$ under different rate $R$, we illustrate rate-perception-classification relationship as shown in Fig. \ref{RDC-fig}(b). The figure shows that points with the same color (indicating the same rate $R$) are about parallel to the perception-axis, which coincides with our theoretical result that the rate only depends on the classification constraint and there is no tradeoff with perception constraint. When $R$ increases, we can obtain higher classification accuracy but the increment above $R\geq 8$ is marginal. We can see that under each level of accuracy (or rate), the potential of zero perception loss remains unaffected. 

\begin{figure}[!htbp]
  \center \includegraphics[width=0.65\textwidth]{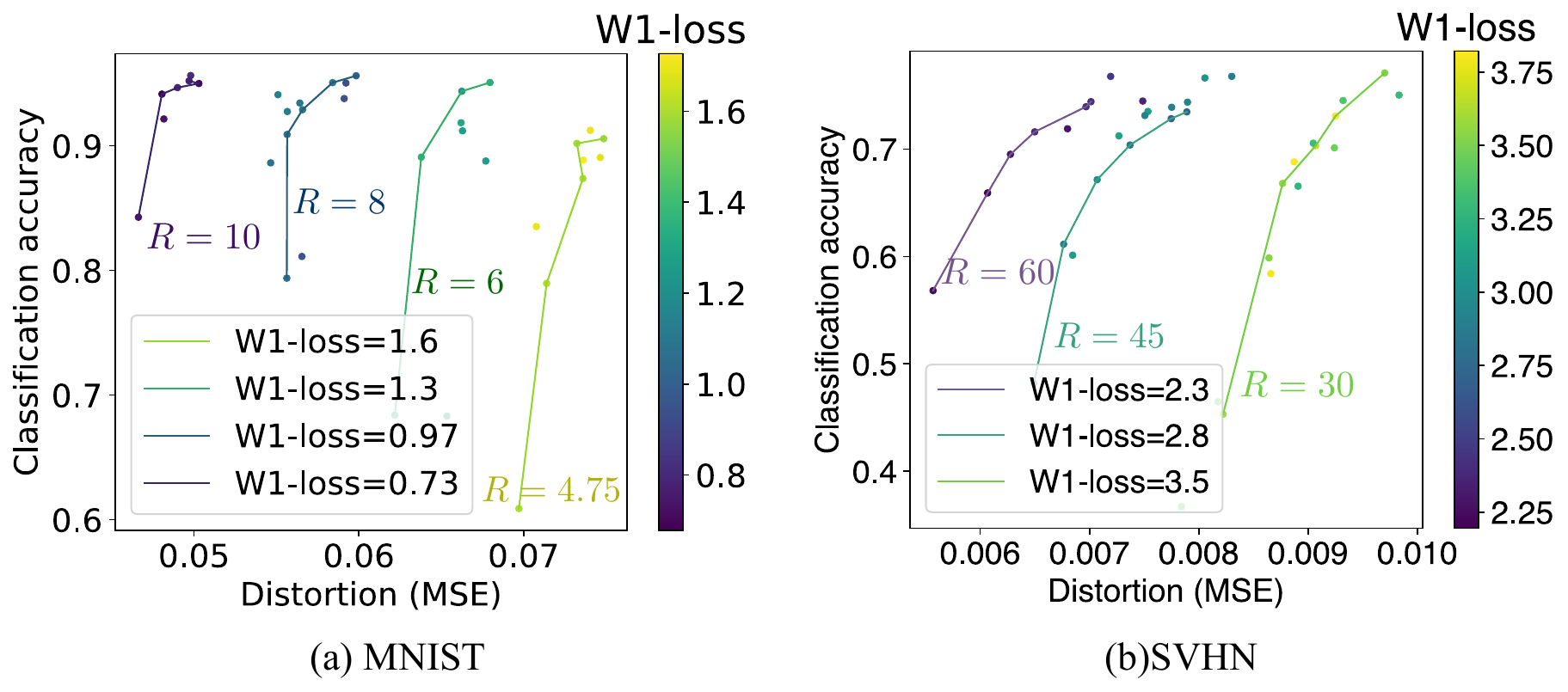}
    \caption{Rate-distortion-classification tradeoffs along different levels of perception quality for (a) MNIST, (b) SVHN datasets. We draw lines to connect the points with approximately the same W1-loss.}\label{RDC-P-fig}
\end{figure}

\begin{figure}[!htbp]
  \center \includegraphics[width=0.65\textwidth]{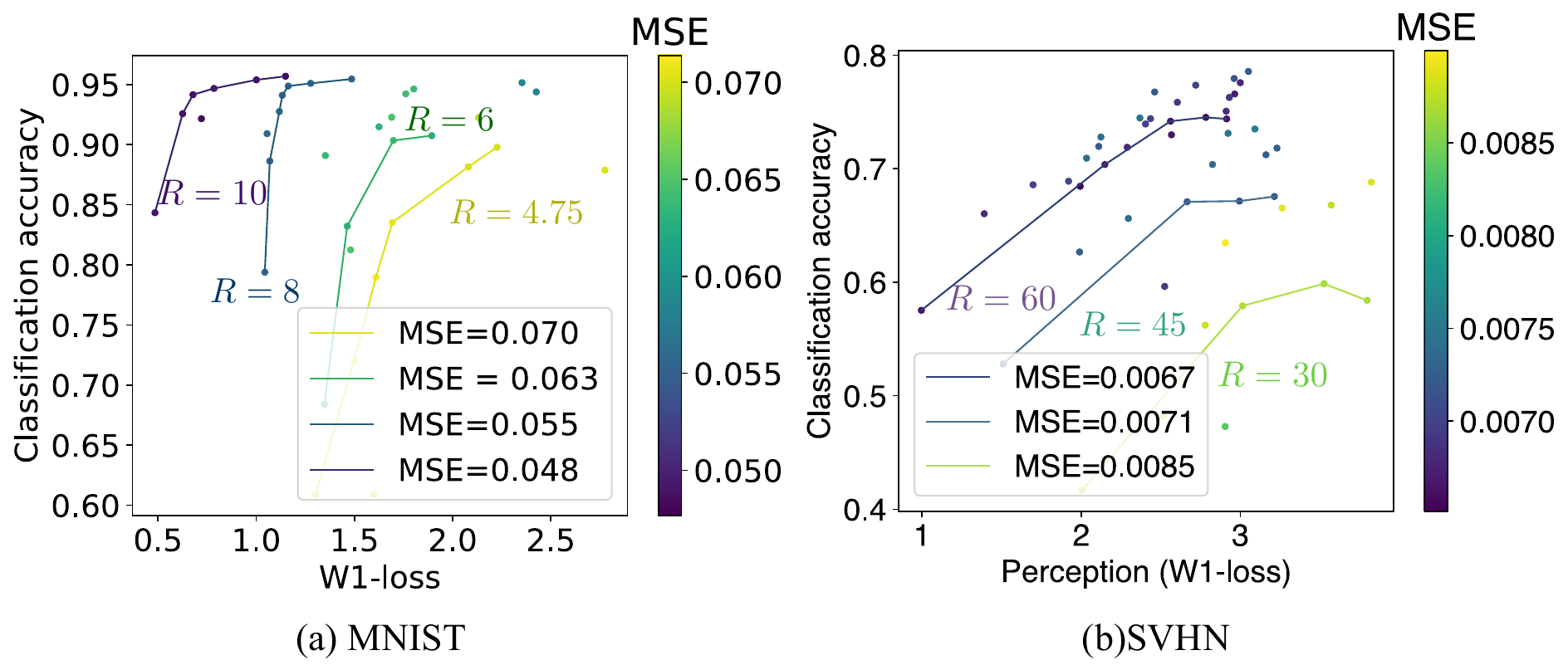}
    \caption{Rate-perception-classification tradeoffs given different levels of distortion for (a) MNIST, (b) SVHN datasets. We draw lines to connect the points with approximately the same MSE.}\label{RPC-D-fig}
\end{figure}

\subsubsection{RDC given P and RPC given D} To jointly consider distortion, perception and classification losses, we train the model with $\lambda_d=1$ and varying values of $\lambda_p$ and $\lambda_c$.

In Fig. \ref{RDC-P-fig}(a) and Fig. \ref{RDC-P-fig}(b), we show the RDP tradeoff given a certain level of perception loss on the MNIST and SVHN datasets respectively, where \emph{we connect the points with approximately the same W1-loss}. For the MNIST dataset, rates are set to be $R=4,75, 6, 8, 10$ respectively, while for the SVHN dataset, rates $R=\text{dim}\times\log_2(L)$ are set with (dim, $L$) pairs $(10,8)$, $(15,8)$, $(20,8)$.

Fig. \ref{RPC-D-fig}(a) and Fig. \ref{RPC-D-fig}(b) depict the RPC tradeoff given different levels of distortion, with the color representing the value of MSE. \emph{We draw lines to connect the points with approximately the same MSE under each rate}, meaning each curve represents the relationship between classification accuracy and perception loss under a specific distortion level (e.g., 0.070, 0.063, 0.055, 0.048 for the MNIST dataset). It is observed that by jointly considering distortion loss and connecting points with the same level of MSE, the tradeoff between perception and classification emerges, which aligns with the results presented in Section \ref{sec-RPC_D}.

\begin{figure}[t]
  \center 
  \includegraphics[width=0.8\linewidth]{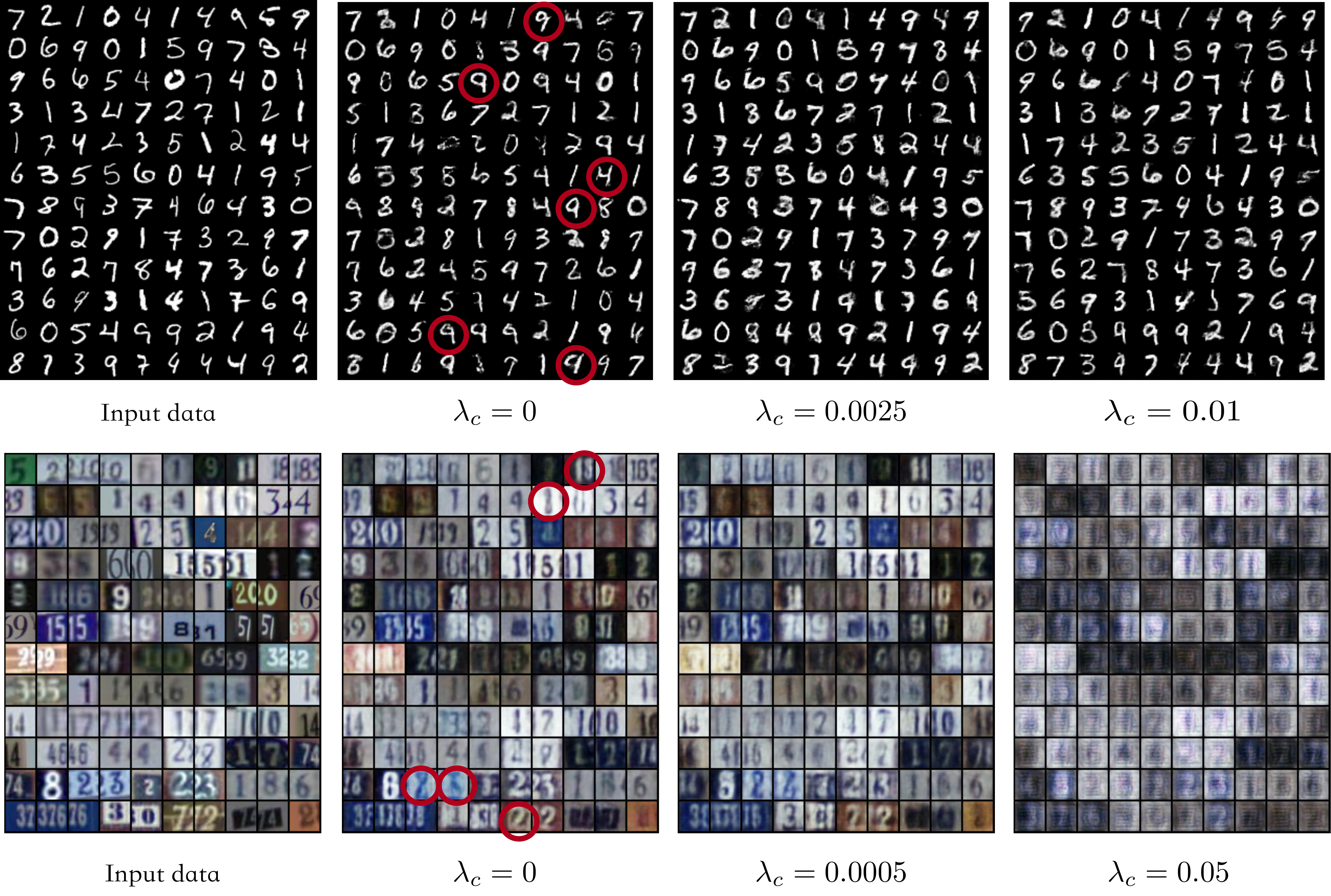}
    \caption{Visual reconstructions for MNIST and SVHN datasets of the decoder with different $\lambda_c$ given $\lambda_p=0.015$ and $R=4.75$ (for MINST) and $\lambda_p=0.00125$ and $R=60$ (for SVHN) respectively.}\label{AccVisual} 
\end{figure}

\subsubsection{Improvement on the accuracy}
From the experimental results, we observe that by taking the classification loss into account in our approach, sacrificing a small amount of distortion or perception quality can lead to a great improvement in classification accuracy compared to the RDP function\cite{RethinkingRDP,UniversalRDPs}. As shown in Table \ref{Acc}, for the MNIST dataset, we set $\lambda_d = 1$, $\lambda_p = 0.015$ and adjust the value of $\lambda_c$ which controls the weight of cross-entropy loss. When $\lambda_c=0$, i.e., the classification is not considered, the accuracy is only $62\%$. However, as we raise $\lambda_c$ to 0.006, the accuracy jumps to $90\%$ with $2\%$ increase on MSE and $1\%$ increase on W1-loss. Similar results can be obtained for the SVHN dataset. As shown in Table \ref{Acc}, when $\lambda_c$ increases from 0 to 0.0015, a 17\% improvement in accuracy is observed, with a cost of 0.44 W1-loss and 0.001 MSE. Meanwhile, by giving classification a much larger weight ($\lambda_c=0.05$) as shown in the last row of Table \ref{Acc}, we can achieve a high accuracy of approximately 0.87, approaching the accuracy tested on the original dataset using the same pre-trained classifier (approximately 0.91).

  \begin{table}[!htbp]
    \center
    \caption{Comparison of distortion, perception and accuracy quality on different $\lambda_c$ for MNIST dataset ($\lambda_d=1$, $\lambda_p=0.015$, $R=4.75$) and SVHN dataset ($\lambda_d=1$, $\lambda_p=0.00125$ and $R=60$).}\label{Acc}
    \begin{tabular}{c|lccc}
      \toprule
      Datasets & $\lambda_c$ & Accuracy   &MSE  &W1-loss    \\ 
      \midrule
      \multicolumn{1}{c|}{\multirow{4}{*}{MNIST}} & 0    &  0.6218  & 0.0732   & 1.2255    \\
      & 0.0025   & 0.7612  & 0.0736   & 1.2412    \\
      & 0.006 & 0.9061  & 0.0747 & 1.2889\\
      & 0.01 & 0.9307 &	0.0753	& 1.4128 \\
      \midrule
      \multicolumn{1}{c|}{\multirow{4}{*}{SVHN}} & 0    &  0.5744  & 0.0061   & 1.1993    \\
      & 0.001 & 0.7189  & 0.0068 & 2.2892\\
      & 0.0015 & 0.7441 &	0.0070	& 2.4351 \\
      & 0.05 & 0.8730	 & 0.0151 &	5.0038\\
      \bottomrule
      \end{tabular}
      \vspace{-0.5cm}
  \end{table}

  Significant improvements can also be observed visually when incorporating the classification constraint. As shown in Fig. \ref{AccVisual}, when $\lambda_c=0$, numerous misclassified samples are present in the images, and it is very common for a input ``4'' to be reconstructed as a ``9'' in MNIST (or get a ``8'' given a input ``2'' and ``3'' in SVHN), as marked by red circles. When the weight of classification increases, such problems disappear and there is no dramatic change on the perceptual quality. However, when the weight of cross entropy loss is set too high (as $\lambda_c=0.05$), the distortion and perceptual quality will be degraded dramatically.

\subsection{Vanishment of Tradeoffs When Rate Increases}
According to the results in Section \ref{SecDistortion}, when there is no constraint on the rate in the lossy compression framework, or equivalently, when the noise power $\sigma_N^2$ is zero, perfect reconstruction is possible and tradeoffs should no longer exist. In this experiment, we set the rate $R$ as $2\times 28\times 28\times \log 256=12544$ (i.e., twice the size of the upper bound of an original image from the MNIST dataset) and train the network using the same weights of $\lambda_d$, $\lambda_p$ and $\lambda_c$ as in the case of small rates.

\begin{figure}[!htpb]
  \center \includegraphics[width=0.85\textwidth]{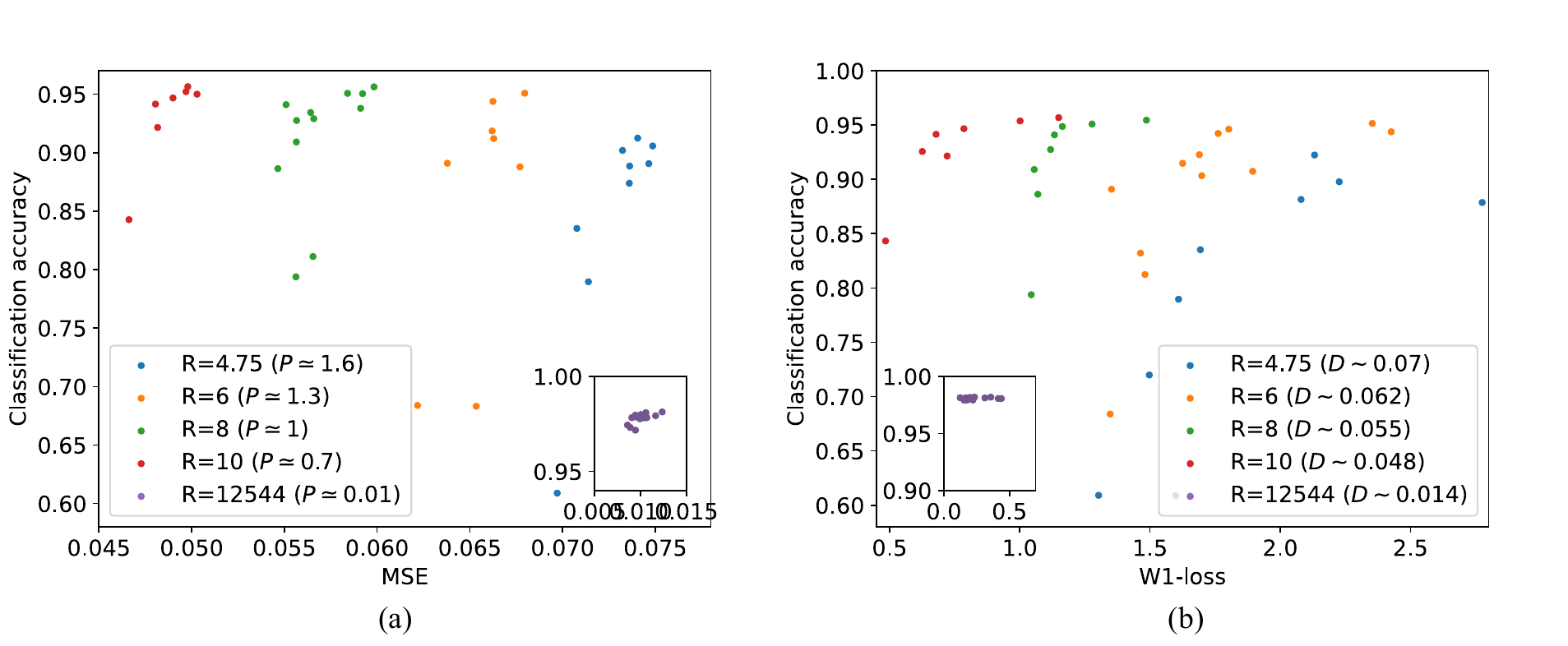}
    \caption{The vanishment of (a) RDC tradeoff given P and (b) RPC tradeoff given D when the rate $R$ increases for MNIST dataset.}\label{RDCPC-largeR-fig}
\end{figure}

From Fig. \ref{RDCPC-largeR-fig}, we observe that as the rate increases, both the $(D, C)$ and $(P, C)$ points become more concentrated. By setting a sufficiently high rate (e.g., $R=12544$ bits), the tradeoffs in RDC and RPC given D are almost gone. In this scenario, the classification accuracies consistently remain at 0.975, regardless of the MSE ranging from 0.006 to 0.014, and the W-1 loss remains consistently below 0.5.

\section{Discussions}
In previous sections, we characterized the properties of RDC and RPC functions for Binary and scalar Gaussian cases. We also explored the decisive role of distortion in the existence of tradeoffs. The above experiments empirically demonstrated the effectiveness of our theoretical results, verifying the characteristics of different tradeoffs on real-world datasets. In the following, we identify several concerns and suggest possible directions for future research to expand upon our framework.

\subsubsection{Joint consideration of rate, distortion, perception, and classification} 
In this paper, we consider two ternary tradeoffs, namely RDC and RPC. The joint consideration of four aspects of rate, distortion, perception, and classification remains unresolved. Note that the techniques involved in the derivation of closed-form RDC and RPC diverge from those of RDP and IB problems, which involve carefully determining the activeness boundary or delicately finding a solution with perfect perception. Consequently, extension to the case with all three constraints \eqref{Distortion}-\eqref{Classification} is not straightforward, even in binary or scalar Gaussian scenarios. A primary attempt may involve studying the boundary at which certain constraints become inactive, which is more complicated than the ternary case. Recent studies have shown that the RDP function and IB problem can be formulated as an optimal transport problem \cite{Chen_IB-OT2023,OT_book}, and then the critical transition can be identified \cite{RDP-OP2024}. Analogous methodologies may be applied to the RDPC function. Meanwhile, techniques from network information theory \cite{Network_IT_book_El_Gamal_Kim_2011} may also be used to characterize the bound of the RDPC function.


\subsubsection{Extension to other source distributions}
The theoretical results of RDC and RPC functions mainly focus on Binary and scalar Gaussian cases, which is a strong foundation. However, it is plausible to derive iterative algorithms to characterize the RDC, RPC, and even RDPC functions for general source distributions. In \cite{IB}, the authors proposed to solve the IB problem with a natural extension of the Blahut-Arimoto (BA) algorithm based on self-consistent equations. In \cite[Section III-A]{Explainable_SemCom_Ma2023}, a modified version of the BA algorithm is also applied to compute the optimal solution of the RDP problem for general sources, suggesting the potential to extend such alternating minimization process to our RDC, RPC, or RDPC problems. However, such an extension is not straightforward since the convergence and correctness of the algorithms need to be further verified. Meanwhile, since RDP and IB have been formulated as optimal transport problems \cite{Chen_IB-OT2023,RDP-OP2024} and can be solved by modified sinkhorn algorithms\cite{OT_book,Sinkhorn}, similar techniques could potentially be applied to tackle our RDC, RPC, and RDPC functions.

\subsubsection{Extension to general distortion and other noise models}
In practice, many deep learning-based methods utilize more general distortion metrics beyond pixel-level MSE. For example, feature reconstruction loss \cite{PerceptualLosses} computes the feature-level MSE, where the features correspond to the outputs of the $j$th layer of a VGG-16 network. By incorporating more generalized distortion and perceptual metrics into the RDPC framework, we can broaden the scope of applications for our theories. Meanwhile, we re-examined the tradeoffs under additive Gaussian noise in Section \ref{sec-comparison_lossy_degra}. In forthcoming studies, we can explore more general degradation scenarios, such as fading channels, to make our theory more applicable to practical wireless communication systems.

\subsubsection{Traversing different tradeoffs in real time}
In real-world communication systems, dynamically navigating tradeoffs in real time is vital to meet diverse task demands. In prior works \cite{PD-tradeoff, RethinkingRDP}, the authors proposed to employ generative adversarial networks (GAN) to traverse the DP tradeoff and the RDP function. However, to achieve different tradeoffs between rate, distortion, and perception, one needs to adjust loss functions and retrain the model, which is resource-consuming and impractical in real-world systems. Recent advances in image denoising\cite{DP-tradeoff_Wasserstein_Freirich2021,PSCGAN} have made progress on traversing DP tradeoff at inference time for specific tasks, e.g., super-resolution or Gaussian deblurring. Nevertheless, the flexible navigation of DP tradeoffs for general inverse problems remains an open challenge, let alone for the RDP and newly proposed RDC and RPC tradeoffs. 



\section{Conclusions}\label{SecConclusion}
In this paper, we integrated data reconstruction tasks and generative tasks into the IB principle, focusing on the lossy compression problem with task-oriented constraints. We first investigated the rate-distortion-classification (RDC)   and the rate-perception-classification (RPC) tradeoffs and derived the closed-form expressions for the RDC and RPC functions in binary and scalar Gaussian cases. These new results complement the IB principle and RDP tradeoff. Meanwhile, our analysis of lossy compression and signal restoration frameworks reveals that the presence of source noise, which can be interpreted as a specific level of distortion in lossy compression, plays a decisive role in the existence of tradeoffs. Furthermore, experiment results on a DL-based image compression justify our theoretical results, providing insights and guidelines on extracting task-relevant information to fulfill diverse objectives.

{\appendices

\section{Proof of Theorem \ref{TheoremBinaryRDC}}\label{Appendix_Proof_RDC_Bin}
Consider the RDC function in \eqref{RDC} with Hamming distance for a Bernoulli source $X$ and a classification variable $S$ with the binary symmetric joint distribution given by $S=X\oplus S_1$ where $S\sim \text{Bern}(a)$ and $S_1\sim \text{Bern}(p_1)$ ($a,p_1\leq 1/2$).

 From the binary symmetric joint distribution, we can obtain that the marginal distribution of $X$ is $\text{Pr}(X=1)= \frac{a-p_1}{1-2p_1}\triangleq b$. Here we always choose $b$ to be less than $\frac{1}{2}$, otherwise we simply assign $\text{Pr}(X=1)=1-b$.
 
Note that by data-processing inequality\cite[Theorem 2.8.1]{cover2012elements} and $S-X-\hat{X}$, we have
 \begin{align*}
  H(S|\hat{X}) \geq H(S|X) = H(X\oplus S_1|X) = H(S_1).
 \end{align*}
 Thus, when $C<H(S_1)$, the classification constraint $H(S|\hat{X})\leq C$ would be infeasible. We assume $C\geq H(S_1)$ in the following discussion.

We first find a lower bound of $I(X,\hat{X})$ leveraging the constraints and then show that it is achievable. By Mrs. Gerber's Lemma proved in \cite{MrsGerbers}, when $H(S|\hat{X})\leq C$, we have $H(X|\hat{X})\leq H(C_1)$, where $C_1\triangleq\frac{H^{-1}(C)-p_1}{1-2p_1}$ and $H^{-1}:[0,1]\rightarrow [0,1/2]$ denotes the inverse function of Shannon entropy for probability less than $\frac{1}{2}$. Note that $C_1=\frac{H^{-1}(C)-p_1}{1-2p_1}\leq\frac{\frac{1}{2}-p_1}{1-2p_1}=\frac{1}{2}$.
Therefore, we have
\begin{align}
  I(X;\hat{X})=H(X)-H(X|\hat{X})\geq H(b)-H(C_1). \label{LowerC}
\end{align}

\emph{Case 1:} When $D\leq b$ and $D\leq C_1$, since 
\begin{align*}
  \mathbb{E}[\Delta(X,\hat{X})] 
  &= P(X\neq \hat{X}) = P(X\oplus \hat{X} = 1) \leq D,
\end{align*}
we can obtain
\begin{align}
  I(X;\hat{X}) &= H(X) - H(X|\hat{X})\geq H(b) - H(\hat{X}\oplus X) \geq H(b) - H(D).\label{LowerD}
\end{align}
Since $D\leq C_1\leq\frac{1}{2}$, we have $H(D)\leq H(C_1)$. Combining \eqref{LowerC} and \eqref{LowerD}, we have the lower bound of rate $I(X;\hat{X}) \geq H(b) - H(D)$. This lower bound can be achieved by choosing $(X, \hat{X})$ to have the joint distribution given by BSC with $X=\hat{X}\oplus S_2$ where $S_2\sim \text{Bern}(D)$, since $I(X,\hat{X}) = H(X)-H(X|\hat{X}) = H(b)-H(D)$. Here the classification constraint is also satisfied: By $D<C_1=\frac{H^{-1}(C)-p_1}{1-2p_1}$, we have $p_1*D< H^{-1}(C)$ where $p_1*D=p_1(1-D)+D(1-p_1)$, so $H(S|\hat{X})=H(p_1*D)<C$. Thus, $R(D,C)=H(b)-H(D)$.

\emph{Case 2:} When $D\geq C_1$ and $C_1\leq b$, considering \eqref{LowerC} and \eqref{LowerD}, the lower bound is $I(X;\hat{X}) \geq H(b) - H(C_1)$. This can be achieved by choosing $(X, \hat{X})$ to have the joint distribution given by BSC with $X=\hat{X}\oplus S_2$ where $S_2\sim \text{Bern}(C_1)$.
Then $I(X,\hat{X}) = H(X)-H(X|\hat{X}) = H(b)-H(C_1)$, and the distortion constraint is also satisfied, since $\mathbb{E}[\Delta(x,\hat{x})]=P(X\oplus\hat{X}=1)=C_1\leq D$. Thus, $R(D,C)=H(b)-H(C_1)$.

\emph{Case 3:} When $\min\{D,C_1\}> b$, by letting $\hat{X}=0$ with probability $1$, we have $I(X;\hat{X})=H(b)-H(\hat{X}\oplus X)=H(b)-H(b)=0$. The distortion constraint is satisfied, since $\mathbb{E}[\Delta(x,\hat{x})]=P(X\oplus\hat{X}=1)=b\leq D$. Meanwhile, since $C_1>b$, we have $\frac{H^{-1}(C)-p_1}{1-2p_1}>\frac{a-p_1}{1-2p_1}$ by definition, implying $H^{-1}(C)>a$. Thus, we have $H(S|\hat{X})=H(a)<C$, which means that the classification constraint is satisfied.

In summary, when $C\leq H(S_1)$, the problem is feasible; otherwise the RDC function for the binary case is
\begin{align*}
  R(D,C)= \begin{cases}
    H(b) - H(C_1)~ \text{ for } D\geq C_1 \text{ and } C_1\leq b,\\
    H(b) - H(D)~ \text{ for } D< C_1 \text{ and } D\leq b,\\
    0 \qquad \text{ for } \min\{D,C_1\}> b.
  \end{cases}
\end{align*}

\section{Proof of Theorem \ref{TheoremRDCGS}}\label{Appendix_Proof_RDC_GS}
Consider the RDC problem as in \eqref{RDC} with MSE distortion, where $(X,S)$ are jointly Gaussian variable with covariance $\text{Cov}(X,S)=\theta_1$. Similar to \cite{UniversalRDPs,IBGaussian}, the optimal reconstruction $\hat{X}$ should also be jointly Gaussian with $X$, and the parameters to be optimized are expectation $\mu_{\hat{x}}$, variance $\sigma_{\hat{x}}^2$, and covariance $\text{Cov}(X,\hat{X})=\theta_2$.
  
  By utilizing the formula of differential entropy and mutual information of the (joint) Gaussian variables\cite[Chapter 8]{cover2012elements}, the RDC problem can be represented by
  \begin{align}
    R(D,C) = &\min_{\mu_{\hat{x}},\sigma_{\hat{x},}\theta_2} -\frac{1}{2}\log(1-\frac{\theta_2^2}{\sigma_x^2\sigma_{\hat{x}}^2})\label{RDCGS}\\
      \quad\text{s.t.}~ 
      &(\mu_x-\mu_{\hat{x}})^2 +\sigma_x^2+\sigma_{\hat{x}}^2-2\theta_2 \leq D,\tag{\ref{RDCGS}{a}}\label{RDCGS-D}\\
      &-\frac{1}{2}\log(1-\frac{\theta_1^2}{\sigma_s^2\sigma_x^4}\cdot\frac{\theta_2^2}{\sigma_{\hat{x}}^2}) \geq h(S) - C.\tag{\ref{RDCGS}{b}}\label{RDCGS-C}
  \end{align}
  The first observation is that we can assume $\mu_{\hat{x}}=\mu_{x}$ without loss of optimality, since the objective function \eqref{RDCGS} and the classification constraint \eqref{RDCGS-C} are irrelevant with the value of $\mu_{\hat{x}}$, and the distortion \eqref{RDCGS-D} will be further reduced by choosing $\mu_{\hat{x}}=\mu_{x}$: $(\mu_x-\mu_{\hat{x}})^2 +\sigma_x^2+\sigma_{\hat{x}}^2-2\theta_2\geq\sigma_x^2+\sigma_{\hat{x}}^2-2\theta_2$.

  The second observation is that when $C < \frac{1}{2}\log(1-\frac{\theta_1^2}{\sigma_s^2\sigma_x^2})+h(S)$, the classification constraint is infeasible. To make $-\frac{1}{2}\log(1-\frac{\theta_2^2}{\sigma_x^2\sigma_{\hat{x}}^2})$ in \eqref{RDCGS} meaningful, we have $1-\theta_2^2/({\sigma_x^2\sigma_{\hat{x}}^2})>0$, i.e., $\theta_2^2/\sigma_{\hat{x}}^2 <\sigma_x^2$.
  Then, the mutual information between $S$ and $\hat{X}$ is upper bounded by
  \begin{align*}
    I(S,\hat{X}) = -\frac{1}{2}\log(1\!-\!\frac{\theta_1^2}{\sigma_s^2\sigma_x^4}\!\cdot\!\frac{\theta_2^2}{\sigma_{\hat{x}}^2}) \!\leq\! -\frac{1}{2}\log(1\!-\!\frac{\theta_1^2}{\sigma_s^2\sigma_x^2}),
  \end{align*}
  making the \eqref{RDCGS-C} infeasible if $C < \frac{1}{2}\log(1-\frac{\theta_1^2}{\sigma_s^2\sigma_x^2})+h(S)$. Thus, in the following discussion, we assume $C\geq \frac{1}{2}\log(1-\theta_1/({\sigma_s^2\sigma_x^2}))+h(S)$.

  In the case of \eqref{RDCGS-C} not active, by Shannon's classic rate-distortion theory for a Gaussian source with squared-error distortion\cite[Chapter 10]{cover2012elements}, the optimal rate is $\frac{1}{2}\log\frac{\sigma_s^2}{D}$ for $D\leq \sigma_x^2$, achieved by setting $\sigma_{\hat{x}}^2=\sigma_{x}^2-D$ and $\theta_2=\sigma_{x}^2-D$. For the classification constraint \eqref{RDCGS-C}, it is inactive if
  \begin{align*}
    I(S,\hat{X})
    & = -\frac{1}{2}\log(1-\frac{\theta_1^2(\sigma_x^2-D)}{\sigma_s^2\sigma_x^4})\geq h(S)-C,
  \end{align*}
  which is equivalent to $D \leq \sigma_{x}^2 - \frac{\sigma_s^2\sigma_x^4}{\theta_1^2}(1-e^{-2h(S)+2C})$.
  
  We now show that when $D > \sigma_{x}^2 - \frac{\sigma_s^2\sigma_x^4}{\theta_1^2}(1-e^{2C-2h(S)})$, the distortion constraint \eqref{RDCGS-D} will be inactive. The classification constraint \eqref{RDCGS-C} gives us a lower bound of $\frac{\theta_2^2}{\sigma_{\hat{s}}^2}$:
  \begin{align}
    \frac{\theta_2^2}{\sigma_{\hat{x}}^2} \geq \frac{\sigma_s^2\sigma_x^4}{\theta_1^2}(1-e^{-2h(S)+2C}).\label{incorporatingC}
  \end{align}
  Observing that the objective function \eqref{RDCGS} is an increasing function of $\frac{\theta_2^2}{\sigma_{\hat{x}}^2}$, and by incorporating \eqref{incorporatingC}, we can obtain
  \begin{align*}
    I(X,\hat{X}) 
    \geq -\frac{1}{2}\log(1-\frac{\sigma_s^2\sigma_x^2}{\theta_1^2}(1-e^{-2h(S)+2C})),
  \end{align*}
  with equality holds at $\frac{\theta_2^2}{\sigma_{\hat{x}}^2} = \frac{\sigma_s^2\sigma_x^4}{\theta_1^2}(1-e^{2C-2h(S)})$. Without loss of optimality, by choosing $\sigma_{\hat{x}}^2 = \theta_2 = \sigma_s^2\sigma_x^4(1-e^{-2h(S)+2C})/{\theta_1^2}$ and substituting it to the distortion function, we can obtain 
  \begin{align*}
    \mathbb{E}[(X-\hat{X})^2] = \sigma_x^2+\sigma_{\hat{x}}^2-2\theta_2 = \sigma_{x}^2 - \frac{\sigma_s^2\sigma_x^4}{\theta_1^2}(1-e^{-2h(S)+2C}) < D,
  \end{align*}
  i.e., the distortion constraint are also satisfied.

  When $C>h(S)$ and $D>\sigma_{x}^2$, we can simply choose $\sigma_{\hat{x}} = 0$ (i.e., $\hat{X}$ is a constant). Then we have the rate $I(X,\hat{X})=0$, and all the constraints are satisfied: $\mathbb{E}[(X-\hat{X})^2] = \sigma_x^2 < D$ and $h(S|\hat{X}) = h(S) < C$.
  In summary, if $C < \frac{1}{2}\log(1-\frac{\theta_1}{\sigma_s^2\sigma_x^2})+h(S)$, the problem will be infeasible; otherwise, the rate-distortion-classification function with MSE distortion is given by
  \begin{align*}
    R(D,C)= \begin{cases}
      \frac{1}{2}\log\frac{\sigma_x^2}{D} ~  &\text{ for } D\leq \sigma_x^2(1-\frac{1}{\rho^2}(1-e^{-2h(S)+2C})),\\
      -\frac{1}{2}\log(1-\frac{1}{\rho^2}(1-e^{-2h(S)+2C}))~ &\text{ for } D> \sigma_x^2(1-\frac{1}{\rho^2}(1-e^{-2h(S)+2C})),\\
      0\ \ &\text{ for } C>h(S)\quad \text{and}\quad D>\sigma_x^2,
    \end{cases}
  \end{align*}
  where $\rho = \frac{\theta_1}{\sigma_s\sigma_x}$ is the correlation factor of $X$ and $S$, and $h(\cdot)$ is the differential entropy for a continuous variable.

  \section{Proof of Theorem \ref{TheoremBinaryRPC}}\label{Appendix_Proof_RPC_Bin}
  Recall the RPC function in \eqref{RPC} for a Bernoulli source $X$ and a classification variable $S$ with binary symmetric joint distribution given by $S=X\oplus S_1$ where $S\sim \text{Bern}(a)$ and $S_1\sim \text{Bern}(p_1)$ ($a,p_1\leq\frac{1}{2}$). Similar to the RDC case, if $C<H(S_1)$, by data-processing inequality\cite[Theorem 2.8.1]{cover2012elements}, the problem is infeasible. In the following discussion, we assume $C\geq H(S_1)$. 

  \underline{Converse:} By Mrs. Gerber's Lemma proved in \cite{MrsGerbers}, when $H(S|\hat{X})\leq C$, we have $H(X|\hat{X})\leq H(C_1)$, where $C_1\triangleq\frac{H^{-1}(C)-p_1}{1-2p_1}$. Note that $C_1=\frac{H^{-1}(C)-p_1}{1-2p_1}\leq\frac{\frac{1}{2}-p_1}{1-2p_1}=\frac{1}{2}$.
  Therefore, we have $I(X;\hat{X})=H(X)-H(X|\hat{X})\geq H(b)-H(C_1)$.
  
  \underline{Achievability:} To achieve the optimal rate, we need to find the optimal solution $p_a=P(\hat{X}=0|X=0)$ and $p_b=P(\hat{X}=0|X=1)$. From the BSC assumption, we can obtained the marginal distribution of $X$ is $P(X=1)= (a-p_1)/(1-2p_1)\triangleq b$.

  Take the perception constraint to be the total-variation (TV) divergence, i.e.,  
    \begin{align*}
      d_{TV}(p_X,p_{\hat{X}})
      &=|-bp_b+(1-b)(1-p_a)|. 
    \end{align*}
  By the definition of conditional entropy and $S-X-\hat{X}$, we also express $H(S|\hat{X})$ as a function of $p_a,p_b$, and denote $f(p_a,p_b)\triangleq H(S|\hat{X})$.
  
  Consider the solution
  \begin{align}
    &p_b = \frac{(1-b)(1-p_a)}{b},~p_a = g^{-1}(C),\label{sol_p_a}
  \end{align}
  where $g^{-1}:[0,1]\rightarrow[\frac{1}{2},1]$ denote the inverse function of 
    $g(p_a)\triangleq f(p_a,(1-b)(1-p_a)/b)$.
  
  First we show that for any $H(S_1)\leq C\leq H(S)$, we can always find a solution $p_a\geq \frac{1}{2}$ to $g(p_a)=C$, i.e., $g^{-1}(C)$ always exists. To simplify the notation, let
  \begin{align*}
    &A = (1-2p_1)(1-b)p_a+p_1(1-b),\\
    &B = (2p_1-1)(1-b)p_a+p_1b+(1-b)(1-2p_1),\\
    &C = (2p_1-1)(1-b)p_a+(1-p_1)(1-b),\\
    &D = (1-2p_1)(1-b)p_a+(1-p_1)b+(1-b)(2p_1-1).
  \end{align*}
  Note that $A+C=1-b$, $B+D=b$. Then we have,
  \begin{align*}
    g(p_a) &= -A\log\frac{A}{1-b} \!-\! B\log\frac{B}{b} \!-\! C\log\frac{C}{1-b} \!-\! D\log\frac{D}{b}\\
    &= H(A,B,C,D) - H(b),
  \end{align*}
  where $H(A,B,C,D)$ is the entropy with quaternary probability distribution $\{A,B,C,D\}$, and is also a function of $p_a$.
  
  By taking the derivative of $g(p_a)$ over $p_a$, we have
  \begin{align*}
    \frac{\text{d}g(p_a)}{\text{d}p_a} 
    &=(1-2p_1)(1-b)\log\frac{B\cdot C}{A\cdot D}.
  \end{align*}
  By solving $\frac{\text{d}g(p_a)}{\text{d}p_a}=0$, i.e., $B\cdot C= A\cdot D$, we can obtain $p_a=1-b$. When $p_a<1-b$, $\frac{\text{d}g(p_a)}{\text{d}p_a}> 0$, $g(p_a)$ increases; and when $1-b\leq p_a\leq 1$, $\frac{\text{d}g(p_a)}{\text{d}p_a}\leq 0$, $g(p_a)$ decreases. Hence, $g(p_a)$ reaches the maximum when $p_a=1-b$ and 
  \begin{align*}
    g(1-b) = H(b*p_1) + H(b) - H(b) = H(S).
  \end{align*}
  Meanwhile, since 
  \begin{align*}
    g(1)& \!=\! H\big((1-p_1)(1-b),p_1b,p_1(1-b),(1-p_1)b\big)\!-\!H(b)\\
        & \!=\! H(p_1)+H(b)-H(b)= H(p_1),
  \end{align*}
 and $g(p_a)$ continuously decreases from $p_a=1-b$ to $p_a=1$, we can always find a solution $p_a\geq 1-b\geq\frac{1}{2}$ satisfying $g(p_a)=C$ as long as $H(p_1)\leq C\leq H(S)$. Hence the value assignment of \eqref{sol_p_a} is valid.
  
  Then we can see that the pair of $(p_a, p_b)$ defined in \eqref{sol_p_a} makes the perception constraint \eqref{RPC-P} always inactive, since for $p_b=\frac{(1-b)(1-p_a)}{b}$, we have
  \begin{align*}
    d_{TV}(p_X,p_{\hat{X}})=|-bp_b+(1-b)(1-p_a)|=0<P. 
  \end{align*}
  At the same time, we have the classification constraint $H(S|\hat{X})=f(p_a,(1-b)(1-p_a)/b)\!=g(p_a)=C$ for the choice of $p_a=g^{-1}(C)$.  

  By Mrs. Gerber's Lemma\cite{MrsGerbers}, when $H(S|\hat{X})\leq C$, we have $H(X|\hat{X})\leq H(C_1)$, with $C_1$ defined as $C_1=\frac{H^{-1}(C)-p_1}{1-2p_1}$. The equality $H(X|\hat{X})= H(C_1)$ holds when $H(S|\hat{X})=C$. Hence, such choice of $(p_a,p_b)$ achieves the lower bound of $I(X;\hat{X})=H(X)-H(C_1)$ with the perception constraint always inactive.

  \section{Proof of Theorem \ref{TheoremRPCGS}}\label{Appendix_Proof_RPC_GS}
  Consider the RPC function in \eqref{RPC} where $(X,S)$ are jointly Gaussian variable with covariance $\text{Cov}(X,S)=\theta_1$. Similar to \cite{UniversalRDPs,IBGaussian}, the optimal reconstruction $\hat{X}$ should also be jointly Gaussian with $X$, thus the parameters to be optimized are expectation $\mu_{\hat{x}}$, variance $\sigma_{\hat{x}}^2$, and covariance $\text{Cov}(X,\hat{X})=\theta_2$.

  The KL-divergence of two Gaussian varibales $X\sim\mathcal{N}(\mu_x,\sigma^2_{x})$ and $\hat{X}\sim\mathcal{N}(\mu_{\hat{x}},\sigma^2_{\hat{x}})$ is
    $$d_{\text{KL}}(p_X,p_{\hat{X}}) 
    =\frac{1}{2}\log\frac{\sigma_{\hat{x}}^2}{\sigma_{x}^2}+\frac{(\mu_x-\mu_{\hat{x}})^2}{2\sigma_{\hat{x}}^2}+\frac{\sigma_x^2-\sigma_{\hat{x}}^2}{2\sigma_{\hat{x}}^2},$$
where we take the natural base of logarithm.

  By utilizing the formula of differential entropy and mutual information of the (joint) Gaussian variables\cite[Chapter 8]{cover2012elements}, as well as the KL divergence of two Gaussian varibales, the RPC problem can be represented by
  \begin{align}
    R(P,C) = &\min_{\mu_{\hat{x}},\sigma_{\hat{x},}\theta_2} -\frac{1}{2}\log(1-\frac{\theta_2^2}{\sigma_x^2\sigma_{\hat{x}}^2})\label{RPCGS}\\
      \quad\text{s.t.}~ 
      &\frac{1}{2}\log\frac{\sigma_{\hat{x}}^2}{\sigma_{x}^2}+\frac{(\mu_x-\mu_{\hat{x}})^2}{2\sigma_{\hat{x}}^2}+\frac{\sigma_x^2-\sigma_{\hat{x}}^2}{2\sigma_{\hat{x}}^2} \leq P,\tag{\ref{RPCGS}{a}}\label{RPCGS-P}\\
      &-\frac{1}{2}\log(1-\frac{\theta_1^2}{\sigma_s^2\sigma_x^4}\cdot\frac{\theta_2^2}{\sigma_{\hat{x}}^2}) \geq h(S) - C.\tag{\ref{RPCGS}{b}}\label{RPCGS-C}
  \end{align}

  First, it is observed that we can set $\mu_{\hat{x}}=\mu_x$ without any effect of the objective function and the constraint \eqref{RPCGS-C}, while reducing the perception constraint \eqref{RPCGS-P}.

  \underline{Converse:}  Similar to the RDC case, by \eqref{RPCGS-C} we have
    $$\frac{\theta_2^2}{\sigma_{\hat{x}}^2} \geq \frac{\sigma_s^2\sigma_x^4}{\theta_1^2}(1-e^{-2h(S)+2C}),$$
  which gives us a lower bound of the objective function:
   $$I(X;\hat{X})-\frac{1}{2}\log\big(1-\frac{\sigma_s^2\sigma_x^2}{\theta_1^2}(1-e^{-2h(S)+2C})\big),$$
  since $-\frac{1}{2}\log(1-\frac{\theta_2^2}{\sigma_x^2\sigma_{\hat{x}}^2})$ is an increasing function of $\frac{\theta_2^2}{\sigma^2_{\hat{x}}}$. The lower bound is achieved when $\frac{\theta_2^2}{\sigma_{\hat{x}}^2}=\frac{\sigma_s^2\sigma_x^4}{\theta_1^2}(1-e^{-2h(S)+2C})$.

  \underline{Achievability:} Now we show that the lower bound could be achieved with the constraint \eqref{RPCGS-P} always inactive.

  Consider the solution
  \begin{align}
    &\sigma_{\hat{x}}=\sigma_{x},~\theta_2^2 = \frac{\sigma_s^2\sigma_x^6}{\theta_1^2}(1-e^{-2h(S)+2C}).\label{RPC-inactive}
  \end{align}
  Since $\frac{\theta_2^2}{\sigma_{\hat{x}}^2}=\frac{\sigma_s^2\sigma_x^4}{\theta_1^2}(1-e^{-2h(S)+2C})$, we have 
  \begin{align*}
    I(X;\hat{X})&=-\frac{1}{2}\log(1-\frac{\theta_2^2}{\sigma_x^2\sigma_{\hat{x}}^2})=-\frac{1}{2}\log\big(1-\frac{\sigma_2\sigma_x^2}{\theta_1^2}(1-e^{-2h(S)+2C})\big)
  \end{align*}
  achieving the lower bound. Meanwhile, we have $\frac{1}{2}\log\frac{\sigma_{\hat{x}}^2}{\sigma_{x}^2}+\frac{(\mu_x-\mu_{\hat{x}})^2}{2\sigma_{\hat{x}}^2}+\frac{\sigma_x^2-\sigma_{\hat{x}}^2}{2\sigma_{\hat{x}}^2}=0<P$ since $\sigma_{\hat{x}}=\sigma_{x}$ and $\mu_{\hat{x}}=\mu_x$.
  
  Now we verify that the solution \eqref{RPC-inactive} is valid. Since the correlation $\rho = \frac{\theta_2}{\sigma_x\sigma_{\hat{x}}}$ should have value between -1 and 1, we require that
    $\rho^2 = \frac{\theta_2^2}{\sigma^2_x\sigma^2_{\hat{x}}} = \frac{\sigma_2\sigma_x^2}{\theta_1^2}(1-e^{-2h(S)+2C}) \leq 1,$
  which gives us 
    $C\geq \frac{1}{2}\log\Big(1-\frac{\theta_2^2}{\sigma_s^2\sigma_x^2}\Big) + h(S)$.

    In the RDC Gaussian case, we have proved that that when $C < \frac{1}{2}\log(1-\frac{\theta_1}{\sigma_s^2\sigma_x^2})+h(S)$, the classification constraint is infeasible. Thus we can safely assume $C\geq \frac{1}{2}\log\Big(1-\frac{\theta_2^2}{\sigma_s^2\sigma_x^2}\Big) + h(S)$, such that the solution \eqref{RPC-inactive} is valid.

  Hence, \eqref{RPCGS-P} is always inactive and the optimal value is $I(X;\hat{X})=-\frac{1}{2}\log\big(1-\frac{\sigma_s^2\sigma_x^2}{\theta_1^2}(1-e^{-2h(S)+2C})\big)$ when $\frac{1}{2}\log(1\!-\!\rho^2)\!+\!h(S)\leq C\leq h(S)$, solely depending on $C$.


\section{Convexity of RDC function}\label{Appendix_RDC_Convex}
Let $\Delta_n$ denote the simplex of probability $n$-vector. For a vector $\mathbf{q}$ and a matrix $\mathbf{T}$, let $\mathbf{q}[i]$ be the $i$-th entry of $\mathbf{q}$ and $\mathbf{T}[ji]$ be the $j$-th row and $i$-th column of $\mathbf{T}$.  Following the method in \cite{Conditional_Entropy_bound_Wyner_1975}, we first give a geometric interpretation of $R(D,C)$ and then prove the convexity.

Consider discrete random variables $(X, S)$ with joint PMF given by $\mathbf{T}\mathbf{q}$ where $\mathbf{q}\in\Delta_n$, $\mathbf{q}[i]=P(X=i)$ and $\mathbf{T}$ is a $m\times n$ matrix, $\mathbf{T}[ji]=P(S=j|X=i)$ with $i\in\{1,\cdots,n\}$ and $j\in\{1,\cdots,m\}$. 

For $\mathbf{w}\in\Delta_k$, let $\mathbf{w}[\alpha] = P(\hat X=\alpha)$ for $\alpha\in\{1,2,\cdots, k\}$. Let $\mathbf{B}$ be a transition matrix with size $n\times k$, where $\mathbf B=(\mathbf{p}_1, \cdots, \mathbf{p}_k)$ and $\mathbf{p}_{\alpha}\in\Delta_n$ for $\alpha\in\{1,2,\cdots, k\}$. Each choice of $\mathbf{w}$ and $\mathbf{B}$ yields a variable $X^{'}$ with marginal distribution $\mathbf{p}=\mathbf{B}\mathbf{w}=\sum_{\alpha=1}^k \mathbf{w}[\alpha]\mathbf p_\alpha$. Then $\mathbf{T}\mathbf{p}$ yields a variable $S^{'}$ with marginal distribution $\mathbf{Tp}=\mathbf{TB}\mathbf{w}=\sum_{\alpha=1}^k \mathbf{w}[\alpha]\mathbf{Tp}_\alpha$.

For some distortion function $\Delta(x, \hat x)$, define $d_{i,\alpha}=\Delta(i,\alpha)$, $i\in\{1,2,\cdots, n\}$ and $\alpha\in\{1,2,\cdots, k\}$. Denote the distortion vector given $\hat{X}=\alpha$ as $\mathbf{d}_{\alpha}=(d_{1,\alpha},\cdots,d_{n,\alpha})$.

For any choice of $\mathbf{w}$ and $\mathbf{B}$, we can compute
\begin{align}
  &\mathbf{p}=\sum_{\alpha=1}^k \mathbf{w}[\alpha] {\mathbf p_\alpha},\label{map_p}\\
  &\rho = H(X^{'}|\hat X) = \sum_{\alpha=1}^{k} \mathbf{w}[\alpha] {H(\mathbf{p}_\alpha)},\label{map_R}\\
  &\eta = H(S^{'}|\hat X) = \sum_{\alpha=1}^{k} \mathbf{w}[\alpha] {H(\mathbf{Tp}_\alpha)},\label{map_C}\\
  &\delta = \mathbb E[\Delta(X^{'},\hat X)] = \sum_{\alpha=1}^k \mathbf{w}[\alpha]\mathbb E[\Delta(X^{'},\hat X)|\hat X=\alpha]\notag\\
  &\ \ \ \ \ \ \ \ \ \ \ \ \ \ \ \ \ \ \ \ \ = \sum_{\alpha=1}^k \mathbf{w}[\alpha]\sum_{i=1}^n \mathbf p_{\alpha}[i] d_{i,\alpha}\notag\\
  &\ \ \ \ \ \ \ \ \ \ \ \ \ \ \ \ \ \ \ \ \ = \sum_{\alpha=1}^k \mathbf{w}[\alpha]\cdot {\mathbf{d}_\alpha^\top \mathbf{p}_{\alpha}},\label{map_D}
\end{align}
where $H(\mathbf{p}_\alpha)$ is the entropy function with input as a distribution.

Consider the mapping $\mathbf{p}_{\alpha}\to (\mathbf{p}_{\alpha}, H(\mathbf{p}_\alpha), H(\mathbf{Tp}_\alpha), \mathbf{d}_\alpha^\top \mathbf{p}_{\alpha})$, and let $\mathcal S$ be the set of all points of { $(\mathbf{p}_{\alpha}, H(\mathbf{p}_\alpha), H(\mathbf{Tp}_\alpha), \mathbf{d}_\alpha^\top \mathbf{p}_{\alpha})$} for $\mathbf p_{\alpha}\in\Delta_n$.

\begin{lemma}
  The set of all $(\mathbf p, \rho, \eta, \delta)$ determined by \eqref{map_p}-\eqref{map_D} for all $\mathbf{w}\in\Delta_k$, $\mathbf{p}_{\alpha}\in\Delta_n$, $\alpha\in\{1,\cdots,k\}$ is the convex hull of $\mathcal S$, denoted as $\mathcal C$.
\end{lemma}
\begin{IEEEproof}
  On the one hand, relations \eqref{map_p}-\eqref{map_D} represent
  $(\mathbf p, \rho, \eta, \delta)$ as a convex combination with weights $\mathbf w[\alpha]$ of the $k$ points $(\mathbf{p}_{\alpha}, H(\mathbf{p}_\alpha), H(\mathbf{Tp}_\alpha), \mathbf{d}_\alpha^\top \mathbf{p}_{\alpha})$ all belonging to $\mathcal S$. On the other hand, every point of the convex hull $\mathcal C$ of $\mathcal S$ is a convex combination of a finite number of points of $\mathcal S$, hence it is of the form $(\mathbf p, \rho, \eta, \delta)$.
\end{IEEEproof}
Now consider the problem of 
\begin{align}
  F(D,C)=&\max H(X|\hat X)\label{FDC}\\
  &\text{s.t.}~ H(S|\hat X)=C,\notag\\
  &\ \ \ \ \ \mathbb E[\Delta(X,\hat X)]=D.\notag
\end{align}
With the formulation \eqref{map_p}-\eqref{map_D} and the fixed marginal distribution of $X$ as $\mathbf q$, the problem \eqref{FDC} is equivalent to 
\begin{align*}
  F(D,C)=&\sup \{\rho|(\mathbf q, \rho, C, D)\in\mathcal{C}\}.
\end{align*}

\begin{lemma}\label{F_DC_concave}
  $F(D,C)$ is non-decreasing and concave over $(D,C)$ with points $(D,C)$ satisfying $F(D,C)> -\infty$.
\end{lemma}
\begin{IEEEproof}
  \underline{Concavity:}
  For any pairs of $(D_1, C_1)$ and $(D_2, C_2)$ satisfying $F(D_i,C_i)> -\infty\ \ (i=1,2)$, denote
  \begin{align*}
    \rho_1&\triangleq F(D_1,C_1)=\sup \{\rho|(\mathbf q, \rho, C_1, D_1)\in\mathcal{C}\},\\
    \rho_2&\triangleq F(D_2,C_2)=\sup \{\rho|(\mathbf q, \rho, C_2, D_2)\in\mathcal{C}\},\\
    (D_{\lambda}, C_{\lambda}) &\triangleq \lambda (D_1,C_1) + (1-\lambda) (D_2,C_2),\\
    \rho_{\lambda} &\triangleq \lambda \rho_1 + (1-\lambda) \rho_2.
  \end{align*}
  By the convexity of the convex hull $\mathcal{C}$, we have $(\mathbf q, \rho_{\lambda}, C_{\lambda}, D_{\lambda})\in \mathcal{C}$. Then
  \begin{align*}
    F(D_{\lambda}, C_{\lambda})&=\sup \{\rho|(\mathbf q, \rho, C_{\lambda}, D_{\lambda})\in\mathcal{C}\}\geq \rho_\lambda=\lambda F(D_1,C_1) + (1-\lambda) F(D_2,C_2),
  \end{align*}
  which proves the concavity of $F(D,C)$.

  \underline{Monotonicity:} Since $F(D,C)$ is concave, we can show the monotonicity by showing that for any $0\leq D\leq D_{\max}$ and $0\leq C\leq H(S)$ we have
  \begin{align*}
    F(D,C)\leq F(D_{\max}, H(S)),
  \end{align*}
  where $D_{\max}\triangleq \max_{\alpha}\mathbb E[\Delta(X,\alpha)]$.

  First, choosing $\hat X=\alpha^*\triangleq \arg\max_{\alpha}\mathbb E[\Delta(X,\alpha)]$ yields
  \begin{align*}
    H(X|\hat X)=H(X), ~~H(S|\hat X)=H(S), ~~\mathbb E[\Delta(X,\hat X)]=D_{\max},
  \end{align*}
  which means $F(D_{\max}, H(S))=H(X)$.
  Since $H(X|\hat X)\leq H(X)$, we have
  \begin{align*}
    F(D,C)\leq H(X)=F(D_{max},H(S)).
  \end{align*}
  Thus, by concavity, we can conclude for any $0\leq D\leq D_{\max}$, $0\leq C\leq H(S)$, and $\lambda\in[0,1]$,
  \begin{align*}
    F(\lambda D+&(1-\lambda)D_{\max}, \lambda C+(1-\lambda)H(S))\\ &\geq \lambda F(D, C) + (1-\lambda)F(D_{\max}, H(S)) \geq F(D,C),
  \end{align*}
  which shows the monotonicity.
\end{IEEEproof}
With Lemma \ref{F_DC_concave}, we can relax the constraints in $F(D,C)$ with inequality, i.e.,
\begin{align*}
  F(D,C)=&\max H(X|\hat X)\\
  &\text{s.t.}~ H(S|\hat X)\leq C,\notag\\
  &\ \ \ \ \ \mathbb E[\Delta(X,\hat X)]\leq D.\notag
\end{align*}
Then the convexity of $R(D,C)$ can be derived directly from Lemma \ref{F_DC_concave}. 
\begin{theorem}
  $R(D,C)$ function is convex over $(D,C)$ with points $(D,C)$ satisfying $R(D,C)< +\infty$.
\end{theorem}
\begin{IEEEproof}
  The rate-distortion-classification function $R(D,C)$ is defined by
\begin{align*}
  R(D,C) = &\min I(X;\hat{X})\\
    \quad\text{s.t.}~ 
    &\mathbb{E}[\Delta(X,\hat{X})] \leq D, \\
    &H(S|\hat{X}) \leq C,
\end{align*}
Since $I(X;\hat{X})=H(X)-H(X|\hat{X})$, we have
\begin{align*}
  F(D,C)=H(X)-R(D,C).
\end{align*}
By Lemma \ref{F_DC_concave}, we have $F(\lambda D_1+(1-\lambda)D2, \lambda C_1+(1-\lambda)C_2) \geq \lambda F(D_1, C_1) + (1-\lambda)F(D_2, C_2)$, for $\lambda\in[0,1]$, i.e.,
\begin{align*}
  R(\lambda D_1&+(1-\lambda)D2, \lambda C_1+(1-\lambda)C_2) \leq \lambda R(D_1, C_1) + (1-\lambda)R(D_2, C_2),
\end{align*}
which proves the convexity of $R(D,C)$.
\end{IEEEproof}

  \section{About the Operational Meaning of RDPC with Common Randomness}\label{Appendix_Operational_Meaning}
\subsection{Optimality with strong asymptotical constraints}
Let $\mathbb{N}_0$ be the set of non-negative integers.
Consider the source $X\sim p_X$ over alphabet $\mathcal X$. Let $\{X_n\}_{n=1}^\infty$ with $X_n\overset{i.i.d}{\sim} p_X$ be an i.i.d process with marginal $p_X$. In \cite{CodingTheorem_RDP_Theis2021}, the authors defined the following (Information) rate function (IRF) and asymptotical achievability, and proved the optimality of the IRF.

\begin{Definition}[(Information) rate function \cite{CodingTheorem_RDP_Theis2021}]\label{Def_IRF}
  For a source $X \sim p_X$ and a set of real-valued functions $D_i$ of joint distributions $p_{X,\hat{X}}$, the (information) rate function (IRF) is defined as
  \begin{align*}
    R(\theta) = \inf I(X,\hat X) \quad ~\text{s.t.}\quad  \forall i : D_i[p_{X,\hat X}] \leq \theta_i.
  \end{align*}
\end{Definition}

\begin{Definition}[Asymptotically achievable with common randomness \cite{CodingTheorem_RDP_Theis2021}]\label{Asym_Achievability}
  For a source $X \sim p_X$ and a given set of constraints $D_i[p_{X,\hat X}] \leq \theta_i, \forall i$, we say that a rate $R$ is (asymptotically) achievable if there exists a sequence of stochastic encoders $f_N : \mathcal X^N \times\mathbb R \to \mathbb N_0$, decoders $g_N:\mathbb N_0\times \mathbb R\to \mathcal X_N$, and a shared random variable $U$ by the encoder and decoder with
  \begin{align*}
    K_N = f(\mathbf X^N,U) \text{ and } \hat{\mathbf X}^N = g(K_N,U)
  \end{align*}  
   such that each joint distribution $p_{X_n,\hat X_n} (n = 1, \cdots , N )$ satisfies the constraints $D_i[p_{X_n,\hat X_n}] \leq \theta_i, \forall i$ and
  \begin{align*}
    \lim_{N\to\infty} H(K_N |U)/N\leq R.
  \end{align*}
\end{Definition}

\begin{lemma}[Theorem 3 in \cite{CodingTheorem_RDP_Theis2021}]\label{IRF_lemma}
  Let an arbitrary source $X \sim p_X$ and constraints $D_i[p_{X,\hat X}] \leq \theta_i$ be given. Then $R < \infty$ is achievable if and only if $R \geq R(\theta)$.
\end{lemma}

Since the constraints \eqref{Distortion}-\eqref{Classification} in the RDPC are all functions of the joint distribution $p_{\hat X, X}$, thereby making the information RDPC function a specific case of the information rate function defined in \cite{CodingTheorem_RDP_Theis2021}, the optimality of RDPC function directly follow the Lemma \ref{IRF_lemma}. Specifically, denote the infimum of achievable rate $R_{a}(\theta)=\inf\{R\text{ is asymptotically achievable with common randomness}\}$, we have the following theorem.

\begin{theorem} \label{RDPC_optimal}
  If $R(D,P,C)< \infty$, then $R_a(D,P,C)=R(D,P,C)$. Specifically, the operational constraints in $R_a(D,P,C)$ is
  \begin{align*}
    &E[\Delta(X_n,\hat{X}_n)]\leq D, \quad \forall n\in[1,2,\cdots, N],\\
    &d(p_{X_n},p_{\hat X_n})\leq P, \quad \forall n\in[1,2,\cdots, N],\\
    &H(S_n|\hat X_n)\leq C, \quad \forall n\in[1,2,\cdots, N].
  \end{align*}
\end{theorem}
\begin{IEEEproof}
  By viewing the RDPC function as a IRF function, the proof directly follows the Lemma \ref{IRF_lemma}.
\end{IEEEproof} 

\subsection{Achievability with Weak Asymptotical Constraints and Optimality of the RDC Function}

  The asymptotical achievability defined in \cite{CodingTheorem_RDP_Theis2021} is very strong, since it requires the constraints $D_i[P_{X_n,\hat X_n}] \leq \theta_i, \forall i$ holding for each element $X_n,\hat X_n (\forall n\in\{1,\cdots, N\})$. Usually, we can also define the achievability where the constraints hold in average, i.e, $\frac{1}{N}\sum_{n=1}^N D_i[P_{X_n,\hat X_n}]\leq \theta_i$.

\begin{Definition}[Weak asymptotically achievable with common randomness]
  For a source $(S,X) \sim p_{S,X}$ and given values of $D,P$ and $C$, we say that a rate $R$ is weak (asymptotically) achievable if there exists a sequence of encoders $f_N : \mathcal X^N \times\mathbb R \to \mathbb N_0$, decoders $g_N:\mathbb N_0\times \mathbb R\to \mathcal X_N$, and a shared random variable $U$ by the encoder and decoder with
  \begin{align*}
    K_N = f(\mathbf X^N,U) \text{ and } \hat{\mathbf X}^N = g(K_N,U)
  \end{align*}  
   such that each joint distribution $p_{X_n,\hat X_n} (n = 1, \cdots , N )$ satisfies the constraints
   \begin{align}
    &\frac{1}{N}\sum_{n=1}^NE[\Delta(X_n,\hat{X}_n)]\leq D \label{op_distortion}\\
    &\frac{1}{N}\sum_{n=1}^Nd(p_{X_n},p_{\hat X_n})\leq P,\label{op_perception}\\
    &\frac{1}{N}\sum_{n=1}^NH(S_n|\hat X_n)\leq C,\label{op_classification}
   \end{align}
  and
  \begin{align*}
    \lim_{N\to\infty} H(K_N |U)/N\leq R.
  \end{align*}
\end{Definition}

We denote the infimum of achievable rate $R_{w}(D,P,C)=\inf\{R\text{ is weak asymptotically achievable with common randomness}\}$. Since the strong achievability implies the weak achievability, the RDPC function is still achievable.

In Appendix \ref{Appendix_RDC_Convex}, we have proved the convexity of RDC function. Utilizing its convexity, we can then prove the optimality under the setting of week asymptotical achievability.

\begin{theorem}
  If $R(D,C)< \infty$, then $R_w(D,C)=R(D,C)$.
\end{theorem}

\begin{IEEEproof}
  \underline{Achievability:} If rate $R$ is asymptotically achievable as defined in Definition \ref{Asym_Achievability}, then $R$ is also weak asymptotically achievable. Since $R(D,C)$ is asymptotically achievable, we have $R_w(D,C)\leq R_a(D,C)\leq R(D,C)$.

  \underline{Converse:} The coverse part relies follows the traditional converse proof as in rate distortion theory. For any \emph{weak} achievable $R$, we have
  \begin{align*}
    NR&\geq H(f(\mathbf X^N,U)|U)\\
    &\geq I(\mathbf X^N;f(\mathbf X^N,U)|U)\\
    &\geq I(\mathbf X^N;\hat{\mathbf X}^N|U) \ \ \ \text{[data-processing inequality]}\\
    &=I(\mathbf X^N;\hat{\mathbf X}^N|U) + I(\mathbf X^N;U) \ \ \ \text{[independence of $X$ and $U$]}\\
    &=I(\mathbf X^N;(\hat{\mathbf X}^N,U))\\
    &\geq I(\mathbf X^N;\hat{\mathbf X}^N)\\
    &\geq \sum_{n=1}^N I(X_n;\hat X_n)\\
    &\geq \sum_{n=1}^N R(E[\Delta(X_n,\hat{X}_n)], H(S_n|\hat X_n))\\
    &=n\Big(\frac{1}{n}\sum_{n=1}^N R(E[\Delta(X_n,\hat{X}_n)], H(S_n|\hat X_n))\Big)\\
    &\geq n\Big(R(\frac{1}{n}\sum_{n=1}^NE[\Delta(X_n,\hat{X}_n)], \frac{1}{n}\sum_{n=1}^NH(S_n|\hat X_n))\Big)  \ \ \ \text{[convexity of $R(D,C)$]}\\
    &\geq nR(D,C). \ \ \ \text{[non-increasing monotonicity of $R(D,C)$ over $(D,C)$]}
  \end{align*}
\end{IEEEproof}
}



%
\bibliographystyle{IEEEtran}
\bibliography{refer}

\end{document}